\documentclass[twocolumn,showpacs,preprintnumbers,amsmath,amssymb,aps]{revtex4}
\usepackage{graphicx}
\usepackage{bm}
\usepackage{amstext}

\begin{document}
\preprint{}

\title{A relativistic mean field study of multi-strange system}
\author{M. Ikram$^1$} \email{ikram@iopb.res.in} 
\author{S. K. Singh$^2$} \email{shailesh@iopb.res.in}
\author{A. A. Usmani$^1$} \email{anisul@iucaa.ernet.in} 
\author{S. K. Patra$^2$}\email{patra@iopb.res.in} 
\affiliation{$^1$Department of Physics, Aligarh Muslim University, Aligarh -
202002, India.} 
\affiliation{$^2$Institute of Physics, Sachivalaya Marg, Bhubaneswar -
751005, India.} 

\date{\today}

\begin{abstract}
We study the binding energies, radii, single-particle energies, 
spin-orbit potential and density profile for multi-strange 
hypernuclei in the range of light mass to superheavy region 
within the relativistic mean field (RMF) theory.
The stability of multi-strange hypernuclei as a function of 
introduced hyperons ($\Lambda$ and $\Sigma$) is investigated.
The neutron, lambda and sigma mean potentials are 
presented for light to superheavy hypernuclei.
The inclusion of hyperons affects the nucleon, lambda and sigma 
spin-orbit potentials significantly.
The bubble structure of nuclei and corresponding 
hypernuclei is studied.
The nucleon and lambda halo structure are also investigated.
A large class of bound multi-strange systems formed from the 
combination of nucleons and hyperons 
(n, p, $\Lambda$, $\Sigma^+$ and n, p, $\Lambda$, $\Sigma^-$) 
is suggested in the region of superheavy hypernuclei 
which might be stable against the strong decay. 
These multi-strange systems might be produced in heavy-ion reactions.

\end{abstract}
\maketitle

\section{Introduction}
Hypernuclei provide an opportunity to extend 
our knowledge from normal nucleon-nucleon (NN) interaction to 
hyperon-nucleon (YN) and hyperon-hyperon (YY) interactions.
Many of the single- and few double-lambda hypernuclei 
have been observed experimentally
~\cite{danysz1963,prowse1966,takahashi2001,hashimoto2006}, 
which confirm the existence of S~=~-1 and -2 systems.
Available experimental data is limited for S~=~-2 sector, 
and there is no further information for S~$\geq$~-3 system.
Due to complexity of YY scattering, the production of hypernuclei 
with strangeness beyond S~$\textgreater$~-2 is very difficult 
and not only this, ambiguities also exist in theoretical understanding.
It is well known that the hyperon resides at the centre of 
the nucleus for most of the time, and only two hyperons with 
opposite spin can stay in s-state.
Then further injected hyperons would be sat on p-state would 
have less binding in comparison to s-state and because of this 
the production of S~$\geq$~-3 systems is difficult.
It is obvious, with increasing the strength of strangeness, 
the hypernuclear physics becomes more complicated.
Due to complication and importance of strangeness degree of 
freedom in bound as well as in infinite nuclear system, this 
subject has been draw an attention from last few decades~\cite{
Hiyama2001,Hiyama22001,Hiyama2003,Hiyama2004,Hiyama2010,Hiyama20101,
Hiyama20102,Hiyama2012,Nemura2005,nemura2005,Gal2009,Gal2011,
Randeniya2007,Usmani1995,Usmani2006,Usmani2008,Zhou2008,Vidana2004,
Samanta2008,Samanta2004,Samanta2002,Samanta2006,Gibson1972,Gibson1988,
Dalitz1958,Vretenar1998,Lu2003,gibson1995,Gal2010,Gal2004,hashimoto2006,
rufa1992,schaffner1992,schaffner1993,schaffner1994,schaffner2000,
rufa1987,mares1989,bethe1987,ellis1991,glendennning1992}.

The system containing a variety of multi-strange baryons has a 
unique feature to extend the knowledge on hypernuclear chart with 
strangeness of S~$\geq$~-3 dimension.
Many of the theoretical calculations on multi-strange hadronic system 
have been made which explain the changes and effects occur on bound 
nuclear system due to injection of $\Lambda$ hyperon~\cite{rufa1990,
schaffner1992,rufa1987,mares1989,bethe1987,ellis1991,glendennning1992}.
%The presence of hyperons plays a significant role to make the 
%EOS softer for neutron star~\cite{ellis1991,glendennning1992}.
In early investigations within mean field, Rufa {\it et al.} 
suggested the stability of multi-lambda hypernuclei and also 
discussed the pure lambda matter and lambda droplets~\cite{rufa1990}.
The calculations were performed by considering NN and 
$\Lambda$N interactions as a whole for describing the 
multi-strange system.
Even though, they studied the multi-lambda hypernuclei but 
without including the YY interaction.
And the work was limited for medium-heavy spherical nuclei. 
In present work, we make a complete study of multi-strange 
hypernuclei within RMF formalism incorporating YY interaction 
over the periodic chart from light mass to superheavy nuclei by 
introducing $\Lambda$ as well as $\Sigma$ hyperons.

Informations gathered from multi-strange systems are 
quite useful for studying or simulating the structure of 
highly dense astrophysical objects. 
In such a system, there is a possibility of existence of 
bunch of lambdas which is heavier than nucleons. 
Not only this, the existence of all variety of hyperons 
alongwith nucleons is also possible inside the core of 
neutron star at extreme conditions.
In addition to lambda, the production of sigma hypernuclei 
is more difficult because of $\Sigma-$hyperon has repulsive 
nature in nuclear matter with potential depth of 30 MeV.
But the production of $^4_\Sigma$He~\cite{hayano1989,nagae1998}, 
reflects that the many of others $\Sigma-$hypernuclei might be produced.
%It is to expect that, $\Sigma^-$ appears at lower densities 
%than $\Lambda$ via $e^-+n \rightarrow \Sigma^-+\nu$, even 
%though $\Sigma^-$ is more heavier~\cite{vidana2000}. 
%So, $\Sigma^-$ hyperon instead of $\Lambda$ would be the first 
%strange baryon to appear in the core of neutron star.
The $\Lambda$ and $\Sigma^-$ are appeared at high density around 
ten times of normal nuclear matter densities at saturation~\cite{vidana2000}.
On the other way, as well as to search the $\Lambda-$ and 
$\Sigma-$hypernuclei separately, it would be more interesting 
to look for the bound state of $\Lambda$ and $\Sigma$ 
with nucleons, where $\Lambda N-\Sigma N$ coupling 
will play an important role for binding mechanism.
In this context, the array of stable objects composed of 
n, p, $\Lambda$, $\Xi^0$, $\Xi^-$ baryons with 
very high strangeness content and small net charge 
has been investigated in many Refs.~\cite{schaffner1993,
schaffner1994,schaffner2000} within the relativistic mean field model.
%The maximal binding energy per particle for multi-strange 
%nuclei was predicted to be $E_B/A\approx-21$ MeV with strangeness 
%per baryon $f_s\approx1-1.2$, charge per baryon $f_q\approx0.1-0$ 
%and baryon density is around $2.5-3$ times of normal nuclear 
%density~\cite{schaffner1993}. 
Not only this, pure hyperonic bound system involving $\Lambda$, 
$\Xi^0$, $\Xi^-$ hyperons with A~$\geq$~6 has also been 
suggested~\cite{schaffner1993}.
In current study, we search the bound class of multi-strange 
system by considering $\Lambda$, $\Sigma^+$ and $\Sigma^-$ 
as basic participants in addition to nucleons.
The possibilities of bound states of hyperons and nucleons 
(n, p, $\Lambda$, $\Sigma^+$ and n, p, $\Lambda$, $\Sigma^-$) 
are discussed in this paper in the mass range of superheavy region.
It is to expect that the attractive nature of hyperon-hyperon 
interaction allows to form the bound class of multi-strange system
as well as pure hyperonic matter
~\cite{stoks1999,schaffner2000,schaffner1993,schaffner1994}.
%The theoretical results produced in the support of existence 
%of multi-strange systems have great importance and scenario 
%in relativistic heavy ion reactions as well as in astrophysical sites.
%Many of the experiment facilities are available across 
%the world for example,  
%HypHI at GSI, FINUDA at Frascati in Italy, PANDA at FAIR, 
%KaoS at MAMI C in Mainz and Hyperball at KEK and J-PARC 
%in Japan~\cite{site}, which are in active mode to 
%produce the neutron-rich hypernuclei as well as multi-strange 
%system by heavy ion reactions.
%%The informations collected on neutron-rich hypernuclei and 
%%multi-strange system are quite useful to understand the 
%%nature of neutron stars or other compact objects.
%Hyperons are the first exotic particle to appear in the neutron 
%star where the density is assumed to be around $\rho=2-3\rho_0$; 
%$\rho_0$=normal nuclear density, as predicted by several nuclear models.
%In order to study the existence of hyperon in neutron star 
%matter, the many body calculation has been made by employing 
%the YN interaction~\cite{schulze1998}. 
%It is indicated in Ref.~\cite{vidana2000}, that the $\Sigma^0$ and 
%$\Sigma^+-$baryons do not have self-energies as attractive enough 
%to bind these hyperons to the system. 
Various calculations have been performed to study the 
hyperonic system within the RMF with effective interactions
~\cite{prakash1997,knorren1995,schaffner1996,balberg1997,balberg1999}.
The strong strength of attractive YY interaction leads to 
the formation of a system having nucleons and hyperons or 
pure hyperonic matter inclusion of all hyperons such as 
$\Lambda$, $\Sigma^0$, $\Sigma^+$, $\Sigma^-$, $\Xi^-$ 
and $\Xi^0$ at lower densities~\cite{vidana2000}.
It has to be mention that, the presence of hyperons makes 
the EOS as softer, and the inclusion of strong YY interaction 
leads to a further softening of EOS~\cite{vidana2000}.
Incorporation of YY interaction has an impact to study the bound 
system including hyperons as well as infinite nuclear matter system.

%For describing the multi-strange systems, the strange scalar 
%($\sigma^*$) and vector ($\phi$) meson fields have been 
%introduced into RMF to take care of hyperon-hyperon 
%interaction~\cite{schaffner1994,shen2006,weiss2012}.
%It is worthy to mention that, the self-consistent mean 
%field approaches produce successfully the ground state 
%bulk properties of hypernuclei as well as explain the 
%changes which come due to addition of hyperon(s)
%~\cite{lanskoy1998,harada2005,rufa1987,rufa1990}. 

In present work, our motive is to analyze the bulk properties like, 
binding energies, radii, single-particle energies, spin-orbit 
potential and density profile for multi-$\Lambda$ as well as 
multi-$\Sigma$ hypernuclei by continuous injection of hyperons 
with replacing neutrons. 
The stability of multi-strange system as a function 
of introduced hyperons from light mass to superheavy region 
is discussed. 
The neutron, lambda and sigma mean potentials are investigated 
for light to superheavy hypernuclei.
Nucleon, lambda and sigma spin-orbit potentials are also 
displayed for different cases of injected hyperons. 
The bubble structure of nuclei and their disappearance 
by injection of $\Lambda$'s is studied.
On viewing the density profile, the nucleon and lambda 
halo nature are reported.
The bound class of strange and nonstrange baryons 
(n, p, $\Lambda$, $\Sigma^+$ and n, p, $\Lambda$, $\Sigma^-$) 
in the mass range of superheavy hypernuclei is predicted 
which might be produced in heavy ion reactions. 
The paper is organized as follow: the formalism of RMF 
model including YY interaction is given in section 2.  
The results are displayed in section 3.
Paper is summarized in section 4.

%%%%%%%%%%%%%%BEGINGING OF THE FORMALISM %%%%%%%%%%%%%%%%%%%
\section{Formalism }
The structural properties of nuclei are described within the 
framework of effective mean field interactions in relativistic 
and non-relativistic approach. 
RMF model takes care the spin-orbit interaction 
naturally and produces quite remarkable result over the whole periodic 
table including superheavy region~\cite{walecka1974,walecka1986,
serot1992,ring1996,koepf1991,patra2006,skpatra2006,patra2007}.
However, the results produced by original Walecka model was enough 
qualitatively, but there had some modification by Boguta and Bodmer 
to match the results with experimental data in quantitative 
way~\cite{walecka1974,bodmer1977}.
This implies that, for a better understanding of nuclear structure 
studies, it is imperative to include all the possible interactions 
which affect all the physical observables are being to be calculated.
%The relativistic mean field model for nonstrange nuclei 
%has been discussed in many Refs.~\cite{walecka1986,serot1992,
%ring1996,stoker1997,patra1991,patra1993,skpatra1991}.
%The relativistic mean field Lagrangian density for single-$\Lambda$ 
%hypernuclei has been given in many Refs.
%~\cite{rufa1990,glendenning1993,mares1994,
%sugahara1994,vretenar1998,win2008,shen2006,lu2003}. 
To find the results in quantitative way, we include the $\sigma^*$ 
and $\phi$ mesons which simulate the hyperon-hyperon interaction. 
Both relativistic (RMF) and non-relativistic (SHF) mean field 
approaches have been played an interesting as well as successful 
role in order to explore the hypernuclear systems
~\cite{rufa1987,rufa1990,win2011,zhou2008}.
The extension towards multi-strange system has been well 
investigated within RMF however, no experimental data is 
available for such a high strange system of new kinds as 
discussed in Refs.~\cite{rufa1987,rufa1990,schaffner1993,schaffner1994}.
The Lagrangian density for multi-strange hypernuclei is discussed 
in Refs.~\cite{schaffner1994,shen2006}. 
Here, we write the Lagrangian density for multi-strange 
hypernuclei as given below:
\begin{eqnarray} 
\mathcal{L}&=&\mathcal{L}_N+\mathcal{L}_Y+\mathcal{L}_{YY} \;,     Y=\Lambda, \Sigma \;,
\end{eqnarray}
\begin{eqnarray}
{\cal L}_N&=&\bar{\psi_{i}}\{i\gamma^{\mu}
\partial_{\mu}-M\}\psi_{i}
+{\frac12}(\partial^{\mu}\sigma\partial_{\mu}\sigma
-m_{\sigma}^{2}\sigma^{2})                                  \nonumber\\
&-&{\frac13}g_{2}\sigma^{3}                  
-{\frac14}g_{3}\sigma^{4}
-g_{s}\bar{\psi_{i}}\psi_{i}\sigma              
-{\frac14}\Omega^{\mu\nu}\Omega_{\mu\nu}                       \nonumber\\
&+&{\frac12}m_{\omega}^{2}V^{\mu}V_{\mu}              
-g_{\omega}\bar\psi_{i}\gamma^{\mu}\psi_{i}V_{\mu}    
-{\frac14}B^{\mu\nu}B_{\mu\nu}                                  \nonumber \\
&+&{\frac12}m_{\rho}^{2}{\vec{R}^{\mu}}{\vec{R}_{\mu}}  
-{\frac14}F^{\mu\nu}F_{\mu\nu}                        
-g_{\rho}\bar\psi_{i}\gamma^{\mu}\vec{\tau}\psi_{i}\vec{R_{\mu}}  \nonumber \\
&-&e\bar\psi_{i}\gamma^{\mu}\frac{\left(1-\tau_{3i}\right)}{2}\psi_{i}A_{\mu} \;, \nonumber \\ 
\mathcal{L}_{Y}&=&\bar\psi_Y\{i\gamma^\mu\partial_\mu-m_Y\}\psi_Y
-g_{\sigma Y}\bar\psi_Y\psi_Y\sigma                                \nonumber \\
&-&g_{\omega Y}\bar\psi_Y\gamma^{\mu}\psi_Y V_\mu  
+\mathcal{L}_{\rho Y}
+\mathcal{L}_{AY} \;,                                               \nonumber \\ 
\mathcal{L}_{YY}&=&{\frac12}(\partial^{\mu}\sigma^*\partial_{\mu}\sigma^*  
-m_{\sigma^*}^{2}\sigma^{*{2}})
+{\frac12}m_{\phi}^{2}\phi^{\mu}\phi_{\mu}                           \nonumber \\ 
&-&{\frac14}S^{\mu\nu}S_{\mu\nu}       
-g_{\sigma^* Y}\bar\psi_Y\psi_Y\sigma^*	                             \nonumber \\
&-&g_{\phi Y}\bar\psi_Y\gamma^\mu\psi_Y\phi_\mu \;, 
\end{eqnarray}
\begin{equation}
\mathcal{L}_{\rho\Lambda}+\mathcal{L}_{A\Lambda}=0 \;,  
\end{equation}
because of $\Lambda$ is neutral and isoscalar
\begin{equation}
\mathcal{L}_{\rho\Sigma}+\mathcal{L}_{A\Sigma}
=\bar\psi_\Sigma\{g_{\rho\Sigma}\gamma^\mu \vec\tau_\Sigma.\vec R_\mu
+e\frac{(1-\tau_{3\Sigma})}{2}\gamma^\mu A_\mu\}\psi_\Sigma \;,
\end{equation}
here $\psi$ and $\psi_Y$ denote the Dirac spinors for 
nucleon and hyperon, whose masses are M and $m_Y$, respectively.
The quantities $m_{\sigma}$, $m_{\omega}$, $m_{\rho}$ are the 
masses for ${\sigma}-$, ${\omega}-$ ${\rho}-$ mesons. 
The field for the ${\sigma}-$meson is denoted by ${\sigma}$, 
${\omega}-$meson by $V_{\mu}$, ${\rho}-$meson by $R_{\mu}$. 
The quantities $g_s$, $g_{\omega}$, $g_{\rho}$, and $e^2/4{\pi}$=1/137 
are the coupling constants for ${\sigma}-$, ${\omega}-$, ${\rho}-$ 
and photon fields, respectively.
We have $g_2$ and $g_3$ self-interaction coupling constants 
for ${\sigma}-$mesons.
The hyperon-meson coupling constant for strange and non-strange 
mesons are expressed by $g_{\sigma Y}$, $g_{\omega Y}$, 
$g_{\sigma^* Y}$ and $g_{\phi Y}$.
The field tensors of the vector, isovector mesons and of the 
electromagnetic field are given by
\begin{eqnarray} 
\Omega^{\mu\nu}& =& \partial^{\mu} V^{\nu} - \partial^{\nu} V^{\mu} \;,  \nonumber \\
B^{\mu\nu}& =& \partial^{\mu}R^{\nu} - \partial^{\nu}R^{\mu}\;,           \nonumber \\
F^{\mu\nu}& =& \partial^{\mu}A^{\nu} - \partial^{\nu}A^{\mu} \;,           \nonumber \\
S^{\mu\nu}& =& \partial^{\mu}\phi^{\nu} - \partial^{\nu}\phi^{\mu} \;.
\end{eqnarray}
The classical variational principle is used to solve the Lagrangian 
and field equations for hypernuclei are obtained.
The Dirac equation with potential terms for the nucleon is
\begin{equation} 
[-i\alpha.\nabla + \beta(M+S(r))+V(r)]\psi_i=\epsilon_i\psi_i\; ,
\end{equation}
where S(r) is the scalar potential of nucleon written as 
\begin{equation} 
S(r)=g_{\sigma}\sigma(r) \;,
\end{equation}
and V(r) represents the vector potential of nucleon given as 
\begin{equation} 
V(r)=g_{\omega}V_{0}(r)+g_{\rho}\tau_{3}R_{0}(r)+e\frac{(1-\tau_3)}{2}A_0(r)\; ,
\end{equation}
where subscript i = n, p in wavefunction denotes the neutron 
and proton, respectively.
The Dirac equation for $\Lambda-$hyperon is 
\begin{equation} 
[-i\alpha.\nabla + \beta\big(m_\Lambda+S^\Lambda(r)\big)+V^\Lambda(r)]\psi_\Lambda = \epsilon_{\Lambda}\psi_\Lambda \;,
\end{equation}
where $S^\Lambda(r)$ is the scalar potential of $\Lambda-$hyperon given as
\begin{equation} 
S^\Lambda(r)=g_{\sigma\Lambda}\sigma(r)+g_{\sigma^*\Lambda}\sigma^*(r)  \;,
\end{equation}
and $V^\Lambda(r)$ represents the vector potential of $\Lambda-$hyperon written as
\begin{equation} 
V^\Lambda(r)=g_{\omega\Lambda}V_0(r)+g_{\phi\Lambda}\phi(r) \;.
\end{equation}
The Dirac equation for $\Sigma-$hyperon is 
\begin{equation} 
[-i\alpha.\nabla + \beta\big(m_\Sigma+S^\Sigma(r)\big)+V^\Sigma(r)]\psi_\Sigma = \epsilon_{\Sigma}\psi_\Sigma \;,
\end{equation}
where $S^\Sigma(r)$ is the scalar potential of $\Sigma-$hyperon given as
\begin{equation} 
S^\Sigma(r)=g_{\sigma\Sigma}\sigma(r)+g_{\sigma^*\Sigma}\sigma^*(r) \;,
\end{equation}
and $V^\Sigma(r)$ represents the vector potential of $\Sigma-$hyperon written as
\begin{equation} 
V^\Sigma(r)=g_{\omega\Sigma}V_0(r)+g_{\rho\Sigma}\tau_{3\Sigma}R_0(r)+e\frac{(1-\tau_{3\Sigma})}{2}A_0(r)+g_{\phi\Sigma}\phi(r) \;.
\end{equation}
The Klein-Gordon equations for mesons and Coulomb fields are
\begin{eqnarray} 
\{-\bigtriangleup+m^2_\sigma\}\sigma(r) &=& -g_\sigma\rho_s(r)-g_2\sigma^2(r)
-g_3\sigma^3(r) \nonumber \\
&-& g_{\sigma Y}\rho_s^{Y}(r)\; ,                          \nonumber \\ 
\{-\bigtriangleup+m^2_{\sigma^*}\}\sigma^*(r) &=&g_{{\sigma^*} Y}\rho_s^{Y}(r)\; ,  \nonumber \\
\{-\bigtriangleup+m^2_\omega\}V_0(r) &=& g_\omega\rho_v(r)+
g_{\omega Y} \rho_v^{Y}(r)\; ,                               \nonumber \\
\{-\bigtriangleup+m^2_\phi\}\phi_0(r) &=&g_{\phi Y} \rho_v^{Y}(r)\; ,    \nonumber \\
\{-\bigtriangleup+m^2_\rho\}R^0_3(r) &=& g_\rho\rho_3(r)+g_\rho Y\rho_3^Y(r) \;,   Y=\Sigma only \;,  \nonumber \\
-\bigtriangleup A_0(r) &=& e\rho_c(r)+e\rho_c^Y(r)\;,      Y=\Sigma only.
\end{eqnarray}
Here $\rho_s$, $\rho_s^{Y}$ and $\rho_v$, $\rho_v^{Y}$ are 
the scalar and vector density for $\sigma-$ and $\omega-$field 
in nuclear and hypernuclear system which are expressed as
\begin{eqnarray} 
\rho_s(r) &=& \sum_ {i=n,p}\bar\psi_i(r)\psi_i(r)\;,                          \nonumber \\
\rho_s^{Y}(r) &=& \sum_{Y=\Lambda,\Sigma} \bar\psi_{Y}(r)\psi_Y(r)\;,          \nonumber \\
\rho_v(r) &=& \sum_{i=n,p}\psi^\dag_i(r)\psi_i(r) \;,			        \nonumber \\
\rho_v^{Y}(r) &=& \sum_{Y=\Lambda,\Sigma} \psi^\dag_{Y}(r)\psi_Y(r)\;.            
\end{eqnarray}
The vector density $\rho_3(r)$ for $\rho$-field and charge density 
$\rho_c(r)$ for photon field are expressed by 
\begin{eqnarray} 
\rho_3(r) &=& \sum_{i=n,p} \psi^\dag_i(r)\gamma^0\tau_{3i}\psi_i(r)\; ,		        \nonumber \\
\rho_3^Y(r) &=& \sum_{Y=\Sigma} \psi^\dag_Y(r)\gamma^0\tau_{3Y}\psi_Y(r)\; , Y=\Sigma only ,	 \nonumber \\
\rho_c(r) &=& \sum_{i=n,p} \psi^\dag_i(r)\gamma^0\frac{(1-\tau_{3i})}{2}\psi_i(r)\;,      \nonumber \\ 
\rho_c^Y(r) &=& \sum_{Y=\Sigma} \psi^\dag_Y(r)\gamma^0\frac{(1-\tau_{3Y})}{2}\psi_Y(r)\;,   Y=\Sigma only .  
\end{eqnarray}
The various rms radii are defined as
\begin{eqnarray}
\langle r_p^2\rangle &=& \frac{1}{Z}\int r_p^{2}d^{3}r\rho_p\;,                      \nonumber \\
\langle r_n^2\rangle &=& \frac{1}{N}\int r_n^{2}d^{3}r\rho_n\;,                       \nonumber \\
\langle r_m^2\rangle &=& \frac{1}{A}\int r_m^{2}d^{3}r\rho\;,                          \nonumber \\
\langle r_\Lambda^2\rangle &=& \frac{1}{\Lambda}\int r_\Lambda^{2}d^{3}r\rho_\Lambda\;, \nonumber \\ 
\langle r_\Sigma^2\rangle &=& \frac{1}{\Sigma}\int r_\Sigma^{2}d^{3}r\rho_\Sigma\;, 
\end{eqnarray}
for proton, neutron, matter, lambda and sigma rms 
radii, respectively and $\rho_p$, $\rho_n$, $\rho$, 
$\rho_\Lambda$ and $\rho_\Sigma$ are 
their corresponding densities. 
The charge rms radius can be found from the proton rms radius 
using the relation $r_{ch} = \sqrt{r_p^2+0.64}$ by taking into 
consideration the finite size of the proton.
The total energy of the system is given by 
\begin{eqnarray}
E_{total} &=& E_{part}(N, Y)+E_{\sigma}+E_{\omega}+E_{\rho}	\nonumber \\
&+&E_{\sigma^*}+E_{\phi}+E_{c}+E_{pair}+E_{c.m.},   
\end{eqnarray}
where $E_{part}(N, Y)=E_{part}(N, \Lambda, \Sigma)$ is 
the sum of the single particle energies of nucleons 
(N) and hyperons (Y=$\Lambda, \Sigma$).
The other contributions $E_{\sigma}$, $E_{\omega}$, $E_{\rho}$, 
$E_{\sigma^*}$, $E_{\phi}$, $E_{c}$, $E_{pair}$ and 
$E_{cm}$ are from meson fields, Coulomb field, pairing 
energy and the center-of-mass energy, respectively.
For present study, we use NL3* nucleon parameter set 
through out the calculations~\cite{lalazissis09}, which 
produces a good description of nuclear matter as well as 
finite nuclei including superheavy region
~\cite{serot1992,ring1996,patra2007}. 

We adopt the relative $\sigma$ and $\omega$ coupling to 
find the numerical values of hyperon-meson coupling constants.
The ratio of meson-hyperon coupling to nucleon-meson 
coupling is defined by 
$R_\sigma=g_{\sigma Y}/g_s$, $R_\omega=g_{\omega Y}/g_\omega$, 
$R_{\sigma^*}=g_{\sigma^* Y}/g_s$ and $R_\phi=g_{\phi Y}/g_\omega$.
The relative coupling $R_\sigma$, $R_\omega$ for $\Lambda$ 
and $\Sigma$ are adopted from Ref.~\cite{mares1994}.
For meson-hyperon couplings, the naive quark model values 
are used for vector coupling constants.
To incorporate the hyperon-hyperon interaction into the 
calculation, the relative coupling $R_{\sigma^*}$, $R_\phi$ 
are taken from Refs.~\cite{schaffner1994,keil2000,shen2006}.
Here, we consider the coupling strength of sigma-sigma 
interaction same as lambda-lambda interaction as like as used 
by Yang, Shen~\cite{yang2008} and Miyazaki~\cite{miyazaki2004}.
That is, $g_{\phi\Lambda}=g_{\phi\Sigma}=\frac{-\sqrt2}{3}g_{\omega}$ 
from naive quark model and $g_{\sigma^*\Lambda}=g_{\sigma^*\Sigma}=0.69$ 
from Ref.~\cite{schaffner1994}.
In present calculations, to take care of pairing interaction the 
constant gap BCS approximation is used and the centre of mass 
correction is included by the formula $E_{cm}=-(3/4)41A^{-1/3}$.
%%%%%%%%%%%%%%%%%%%%%%%%%%% END OF THE FORMALISM %%%%%%%%%%%%%%%%%%%

%%%%%%%%%%%%%%%%%%%%%%%%%%%%%%%%
\section{RESULTS AND DISCUSSIONS}
\begin{table*}
\renewcommand{\arraystretch}{1.5}
\caption{\label{tab1}Total and single-particle (for s- and p-state) 
binding energies and radii are listed for single-$\Lambda$ hypernuclei. 
The single-particle energies are compared with available experimental 
data. Experimental values are given in 
parenthesis~\cite{hashimoto2006,guleria2012}.}
\begin{ruledtabular}
\begin{tabular}{lccccccccc}
Hypernuclei& BE& $B_\Lambda^s$& $B_\Lambda^p$& $r_{ch}$& $r_t$& $r_p$& $r_n$& $r_\Lambda$ \\
\hline
$^{16}_\Lambda$N   &-130.3091&-14.052(13.76$\pm$0.16)&-3.6920(2.84$\pm$0.16)      &2.562         &2.466         &2.441         &2.509         &2.288\\
$^{16}_\Lambda$O   &-126.7055&-14.055(12.5$\pm$0.35)&-3.6914(2.5$\pm$0.5)      &2.664         &2.474         &2.543         &2.419         &2.288\\
$^{27}_\Lambda$Al  &-228.2085&-19.840&-8.8560      &2.953         &2.812         &2.845         &2.812         &2.324\\
$^{28}_\Lambda$Si  &-236.2911&-20.523(16.6$\pm$0.2)&-9.3438(7.0$\pm$1.0)      &2.990         &2.825         &2.882         &2.799         &2.313\\
$^{32}_\Lambda$S   &-274.1188&-22.147(17.5$\pm$0.5)&-10.455(8.1$\pm$0.6)      &3.157         &2.981         &3.055         &2.942         &2.275\\
$^{40}_\Lambda$Ca  &-347.2458&-20.348(18.7$\pm$1.1)&-11.479(11.0$\pm$0.6)      &3.435         &3.286         &3.342         &3.258         &2.600\\
$^{48}_\Lambda$Ca  &-428.2423&-21.700&-13.380      &3.435         &3.454         &3.349         &3.554         &2.708\\
$^{51}_\Lambda$V   &-456.3979&-22.435(19.97$\pm$0.13)&-14.159(11.28$\pm$0.6)      &3.546         &3.490         &3.460         &3.539         &2.727\\
$^{56}_\Lambda$Fe  &-502.2416&-23.542&-15.265      &3.642         &3.572         &3.557         &3.609         &2.744\\
$^{72}_\Lambda$Ni  &-631.3122&-23.714&-16.640      &3.855         &3.969         &3.782         &4.106         &2.965\\
$^{89}_\Lambda$Y   &-790.4420&-24.543(23.1$\pm$0.5)&-18.095(16.0$\pm$1.0)      &4.216         &4.204         &4.145         &4.269         &3.122\\
$^{90}_\Lambda$Zr  &-797.6272&-24.563&-18.200      &4.243         &4.218         &4.172         &4.275         &3.141\\
$^{132}_\Lambda$Sn &-1121.679&-26.159&-20.958      &4.677         &4.828         &4.618         &4.967         &3.484\\
$^{139}_\Lambda$La &-1184.506&-25.772(24.5$\pm$1.2)&-20.988(20.1$\pm$0.4)      &4.835         &4.895         &4.776         &4.991         &3.624\\
$^{208}_\Lambda$Pb &-1660.121&-26.935(26.3$\pm$0.8)&-23.005(21.3$\pm$0.7)      &5.490         &5.602         &5.439         &5.718         &4.011\\
$^{286}_\Lambda$114&-2097.325&-26.902&-24.024      &6.209         &6.279         &6.163         &6.363         &3.269\\
$^{298}_\Lambda$114&-2169.691&-27.016&-24.143      &6.242         &6.380         &6.198         &6.499         &3.256\\
$^{293}_\Lambda$117&-2132.934&-27.016&-24.143      &6.248         &6.324         &6.203         &6.412         &3.256\\
$^{294}_\Lambda$118&-2134.069&-27.028&-24.173      &6.256         &6.327         &6.210         &6.412         &3.284\\
$^{292}_\Lambda$120&-2108.095&-26.993&-24.213      &6.259         &6.303         &6.213         &6.373         &3.351\\
$^{304}_\Lambda$120&-2190.300&-27.215&-24.316      &6.296         &6.401         &6.252         &6.506         &3.269\\
\end{tabular}
\end{ruledtabular}
\end{table*}

\begin{table*}
\renewcommand{\arraystretch}{1.5}
\caption{\label{tab2}Total and single-particle (for s- and p-state) 
binding energies and radii are listed for single-$\Sigma$ hypernuclei.}
\begin{ruledtabular}
\begin{tabular}{lccccccccc}
Hypernuclei& BE& $B_{\Sigma^+}^s$& $B_{\Sigma^+}^p$& $r_{ch}$& $r_t$& $r_p$& $r_n$& $r_{\Sigma^+}$ \\
\hline
$^{16}_{\Sigma^+}$N   &-130.7513&-13.631&-4.0893      &2.597         &2.465         &2.478         &2.473         &2.300\\
$^{16}_{\Sigma^+}$O   &-124.8292&-11.449&-2.4888      &2.706         &2.488         &2.586         &2.391         &2.363\\
$^{27}_{\Sigma^+}$Al  &-227.6015&-18.490&-8.4594      &2.978         &2.814         &2.872         &2.789         &2.337\\
$^{28}_{\Sigma^+}$Si  &-234.9222&-18.395&-8.2107      &3.017         &2.831         &2.910         &2.779         &2.327\\
$^{32}_{\Sigma^+}$S   &-272.8755&-20.031&-9.6561      &3.187         &2.988         &3.086         &2.921         &2.318\\
$^{40}_{\Sigma^+}$Ca  &-346.4849&-19.008&-10.769      &3.457         &3.288         &3.365         &3.239         &2.594\\
$^{48}_{\Sigma^+}$Ca  &-430.6025&-23.637&-16.267      &3.449         &3.451         &3.364         &3.536         &2.757\\
$^{51}_{\Sigma^+}$V   &-457.6058&-23.251&-15.742      &3.561         &3.488         &3.475         &3.523         &2.746\\
$^{56}_{\Sigma^+}$Fe  &-503.1424&-24.068&-16.321      &3.656         &3.570         &3.572         &3.594         &2.724\\
$^{72}_{\Sigma^+}$Ni  &-636.0653&-28.154&-21.217      &3.866         &3.962         &3.793         &4.089         &2.883\\
$^{89}_{\Sigma^+}$Y   &-792.5601&-26.508&-20.686      &4.226         &4.202         &4.155         &4.258         &3.148\\
$^{90}_{\Sigma^+}$Zr  &-799.4227&-26.162&-20.519      &4.253         &4.217         &4.182         &4.263         &3.192\\
$^{132}_{\Sigma^+}$Sn &-1128.721&-31.481&-26.615      &4.683         &4.823         &4.624         &4.955         &3.455 \\
$^{139}_{\Sigma^+}$La &-1191.296&-29.877&-25.234      &4.841         &4.891         &4.781         &4.981         &3.536\\
$^{208}_{\Sigma^+}$Pb &-1665.060&-32.310&-28.540      &5.494         &5.598         &5.443         &5.709         &3.932\\
$^{286}_{\Sigma^+}$114&-2075.488&-32.596&-29.719      &6.211         &6.275         &6.165         &6.356         &4.471\\
$^{298}_{\Sigma^+}$114&-2148.451&-34.004&-30.666      &6.245         &6.375         &6.200         &6.491         &4.195\\
$^{293}_{\Sigma^+}$117&-2110.368&-32.711&-29.779      &6.250         &6.320         &6.205         &6.405         &4.441\\
$^{294}_{\Sigma^+}$118&-2112.770&-32.602&-29.695      &6.259         &6.324         &6.214         &6.407         &4.459\\
$^{292}_{\Sigma^+}$120&-2087.232&-31.939&-29.210      &6.262         &6.301         &6.217         &6.369         &4.577\\
$^{304}_{\Sigma^+}$120&-2168.554&-33.324&-30.155      &6.299         &6.397         &6.254         &6.498         &4.301\\
\end{tabular}
\end{ruledtabular}
\end{table*}

\begin{table*}
\renewcommand{\arraystretch}{1.5}
\caption{\label{tab3}Total and single-particle (for s- and p-state) 
binding energies and radii are listed for single-$\Sigma$ hypernuclei.} 
\begin{ruledtabular}
\begin{tabular}{lccccccccc}
Hypernuclei&BE& $B_{\Sigma^-}^s$& $B_{\Sigma^-}^p$& $r_{ch}$& $r_t$& $r_p$& $r_n$& $r_{\Sigma^-}$ \\
\hline
$^{16}_{\Sigma^-}$N   &-131.8578&-14.551&-5.1063      &2.524         &2.469         &2.401         &2.545         &2.321\\
$^{16}_{\Sigma^-}$O   &-131.0975&-17.268&-7.2379      &2.618         &2.460         &2.493         &2.450         &2.259\\
$^{27}_{\Sigma^-}$Al  &-233.6156&-24.192&-13.646      &2.918         &2.803         &2.810         &2.831         &2.293\\
$^{28}_{\Sigma^-}$Si  &-242.9184&-26.117&-15.343      &2.954         &2.813         &2.845         &2.815         &2.279\\
$^{32}_{\Sigma^-}$S   &-281.4468&-28.488&-16.885      &3.118         &2.968         &3.015         &2.963         &2.211\\
$^{40}_{\Sigma^-}$Ca  &-355.4514&-27.717&-19.053      &3.405         &3.277         &3.311         &3.274         &2.546\\
$^{48}_{\Sigma^-}$Ca  &-433.1445&-25.891&-17.289      &3.412         &3.452         &3.326         &3.570         &2.581\\
$^{51}_{\Sigma^-}$V   &-463.4586&-28.799&-20.399      &3.523         &3.484         &3.437         &3.552         &2.621\\
$^{56}_{\Sigma^-}$Fe  &-510.6156&-31.242&-23.060      &3.619         &3.566         &3.534         &3.621         &2.667\\
$^{72}_{\Sigma^-}$Ni  &-637.4301&-27.808&-21.023      &3.836         &3.970         &3.762         &4.120         &2.922\\
$^{89}_{\Sigma^-}$Y   &-798.7386&-34.723&-27.787      &4.196         &4.198         &4.125         &4.277         &2.919\\
$^{90}_{\Sigma^-}$Zr  &-806.6424&-35.459&-28.481      &4.223         &4.212         &4.152         &4.282         &2.912\\
$^{132}_{\Sigma^-}$Sn &-1129.498&-34.684&-29.323      &4.661         &4.825         &4.602         &4.973         &3.319\\
$^{139}_{\Sigma^-}$La &-1195.729&-37.227&-32.361      &4.820         &4.891         &4.760         &4.995         &3.474\\
$^{208}_{\Sigma^-}$Pb &-1679.528&-42.618&-38.066      &5.475         &5.596         &5.423         &5.720         &3.626\\
$^{286}_{\Sigma^-}$114&-2096.123&-47.418&-43.950      &6.194         &6.271         &6.148         &6.362         &4.130\\
$^{298}_{\Sigma^-}$114&-2166.315&-46.453&-43.118      &6.228         &6.374         &6.183         &6.499         &4.208\\
$^{293}_{\Sigma^-}$117&-2120.401&-48.304&-44.889      &6.233         &6.316         &6.188         &6.411         &4.161\\
$^{294}_{\Sigma^-}$118&-2123.234&-48.646&-45.247      &6.242         &6.320         &6.197         &6.413         &4.168\\
$^{292}_{\Sigma^-}$120&-2099.320&-49.699&-46.318      &6.246         &6.296         &6.200         &6.373         &4.112\\
$^{304}_{\Sigma^-}$120&-2189.437&-48.468&-45.219      &6.282         &6.395         &6.237         &6.505         &4.250\\
\end{tabular}
\end{ruledtabular}
\end{table*}

\begin{figure}
\vspace{0.6cm}
\includegraphics[width=1.0\columnwidth]{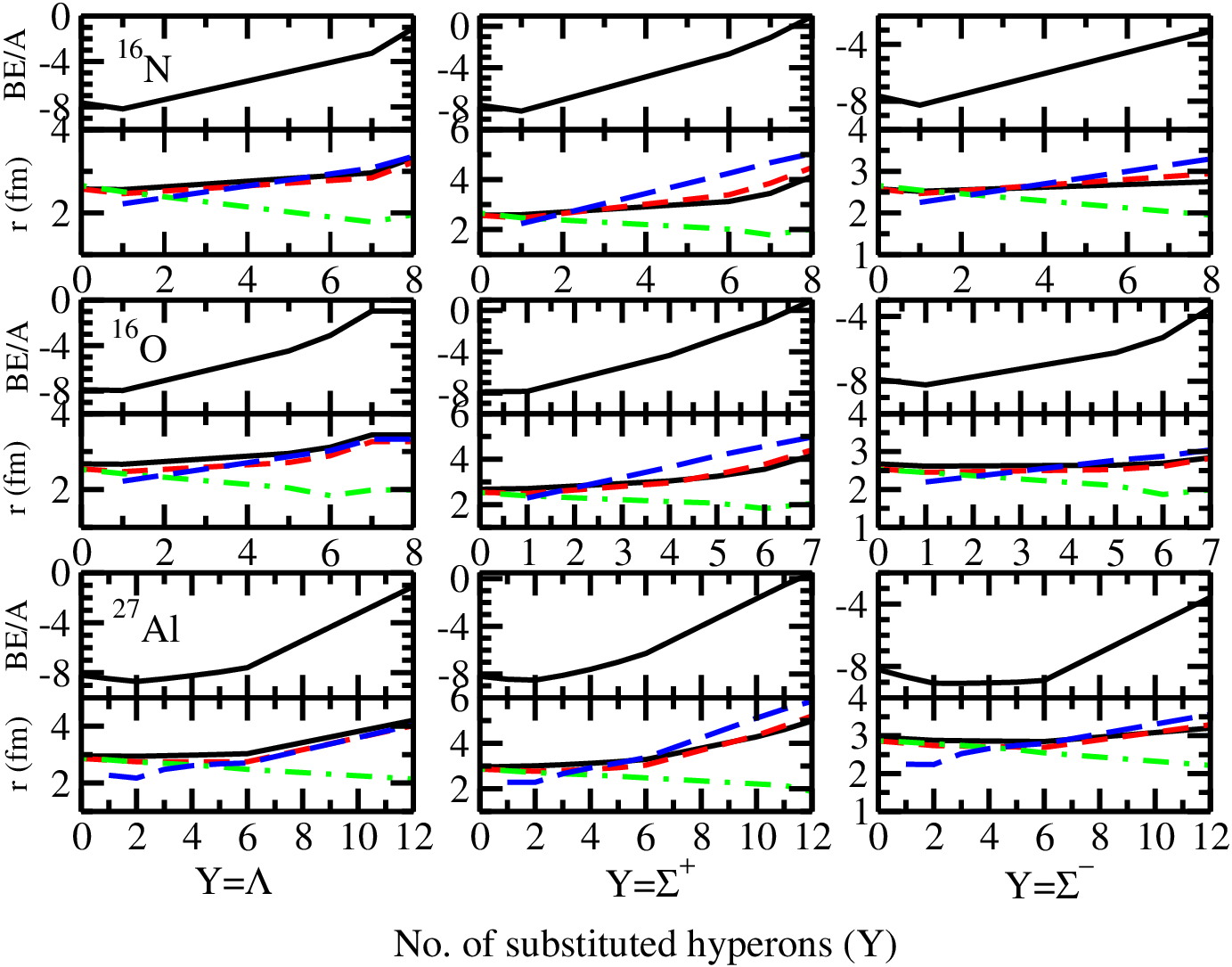}
\caption{\label{fig1} Energy per particle is shown 
as a function of substituted hyperons 
($\Lambda$, $\Sigma^+$, $\Sigma^-$) for 
$^{16}_{nY}$N, $^{16}_{nY}$O and $^{27}_{nY}$Al, where 
Y indicates the injected hyperons ($\Lambda$, $\Sigma^+$, $\Sigma^-$). 
Total ($r_t$), charge ($r_{ch}$), neutron ($r_n$) and 
hyperon ($r_Y$) radii are also displayed as a function of 
substituted hyperons.
Total, charge, neutron and hyperon radii are represented by 
solid, dashed, dot-dashed and long dashed lines with black, 
red, green and blue colors, respectively.}  
\end{figure}

\begin{figure}
\vspace{0.6cm}
\includegraphics[width=1.0\columnwidth]{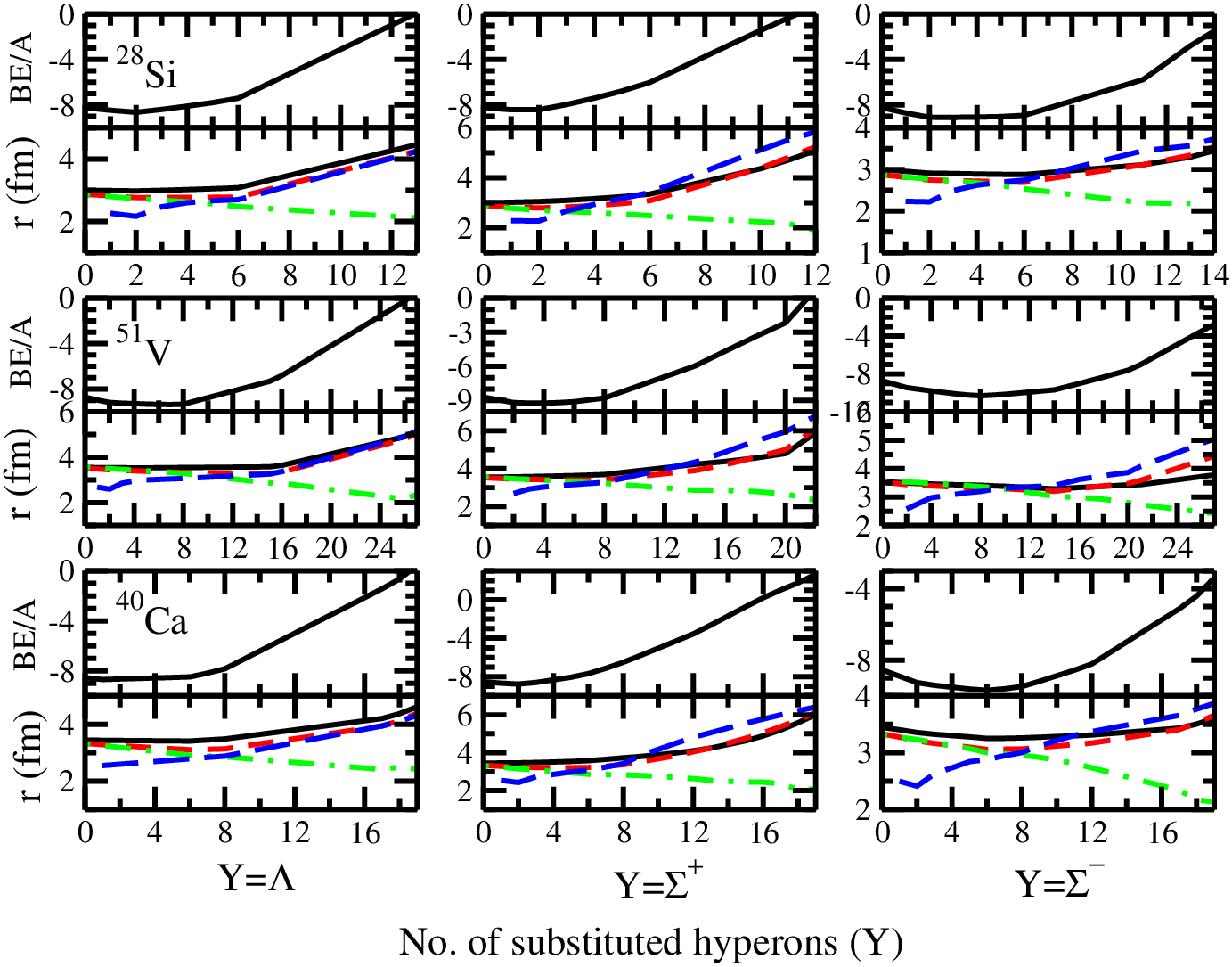}
\caption{\label{fig2} same as Fig. 1 but for $^{28}_{nY}$Si, 
$^{51}_{nY}$V and $^{40}_{nY}$Ca.}  
\end{figure}

\begin{figure}
\vspace{0.6cm}
\includegraphics[width=1.0\columnwidth]{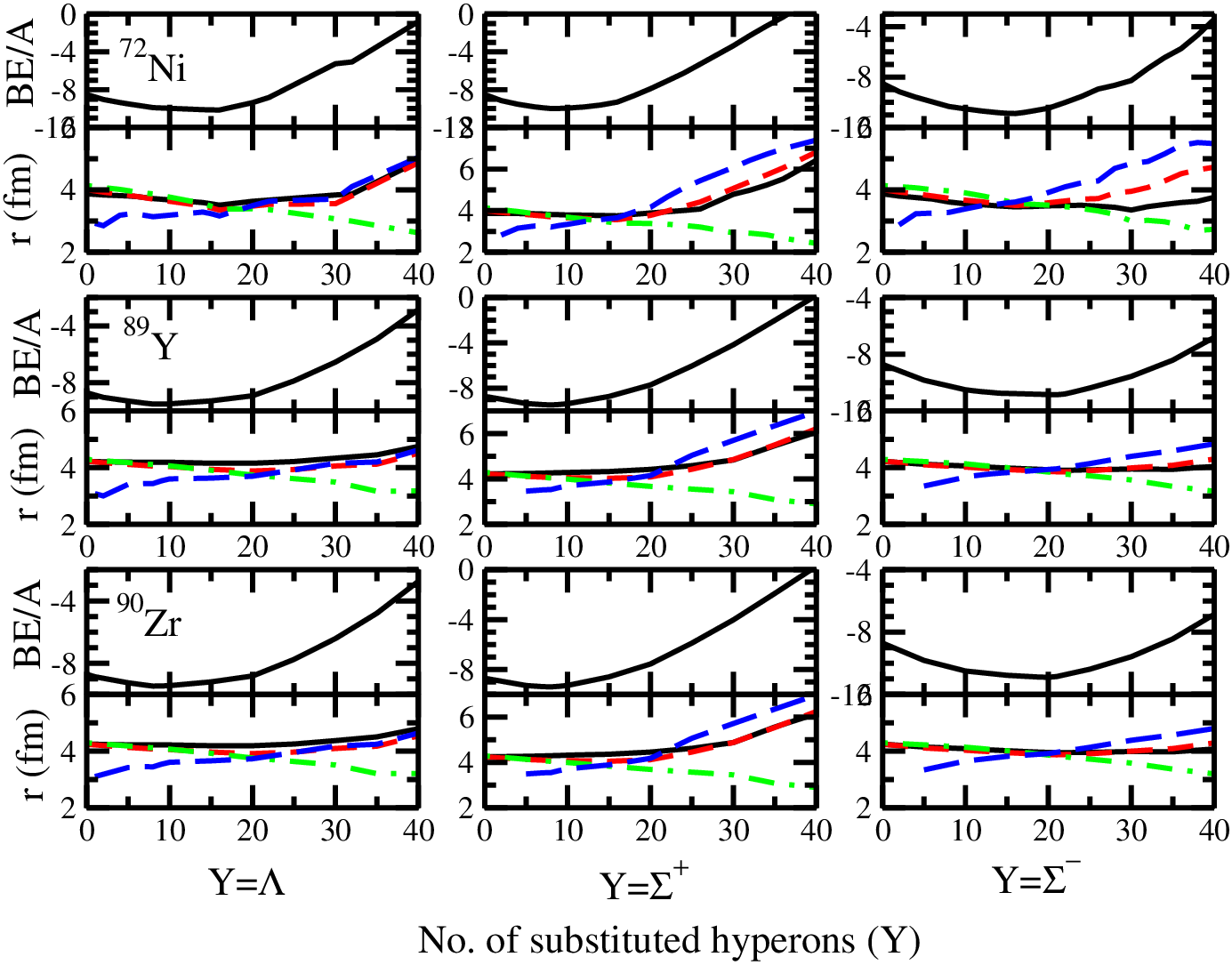}
\caption{\label{fig3} same as Fig. 1 but for $^{72}_{nY}$Ni, 
$^{89}_{nY}$Y and $^{90}_{nY}$Zr.}  
\end{figure}

\begin{figure}
\vspace{0.6cm}
\includegraphics[width=1.0\columnwidth]{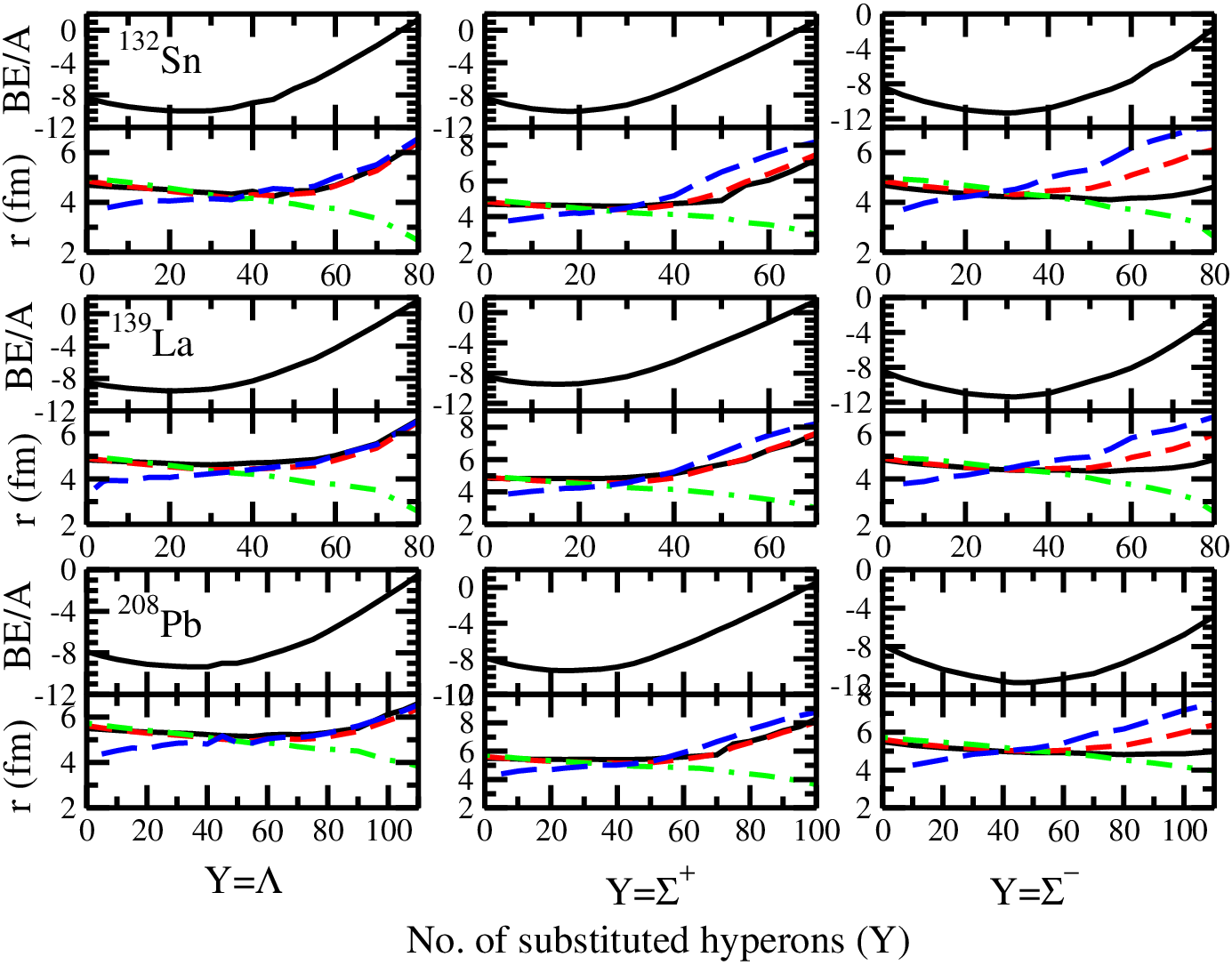}
\caption{\label{fig4} same as Fig. 1 but for $^{132}_{nY}$Sn, 
$^{139}_{nY}$La and $^{208}_{nY}$Pb.}  
\end{figure}

\begin{figure}
\vspace{0.6cm}
\includegraphics[width=1.0\columnwidth]{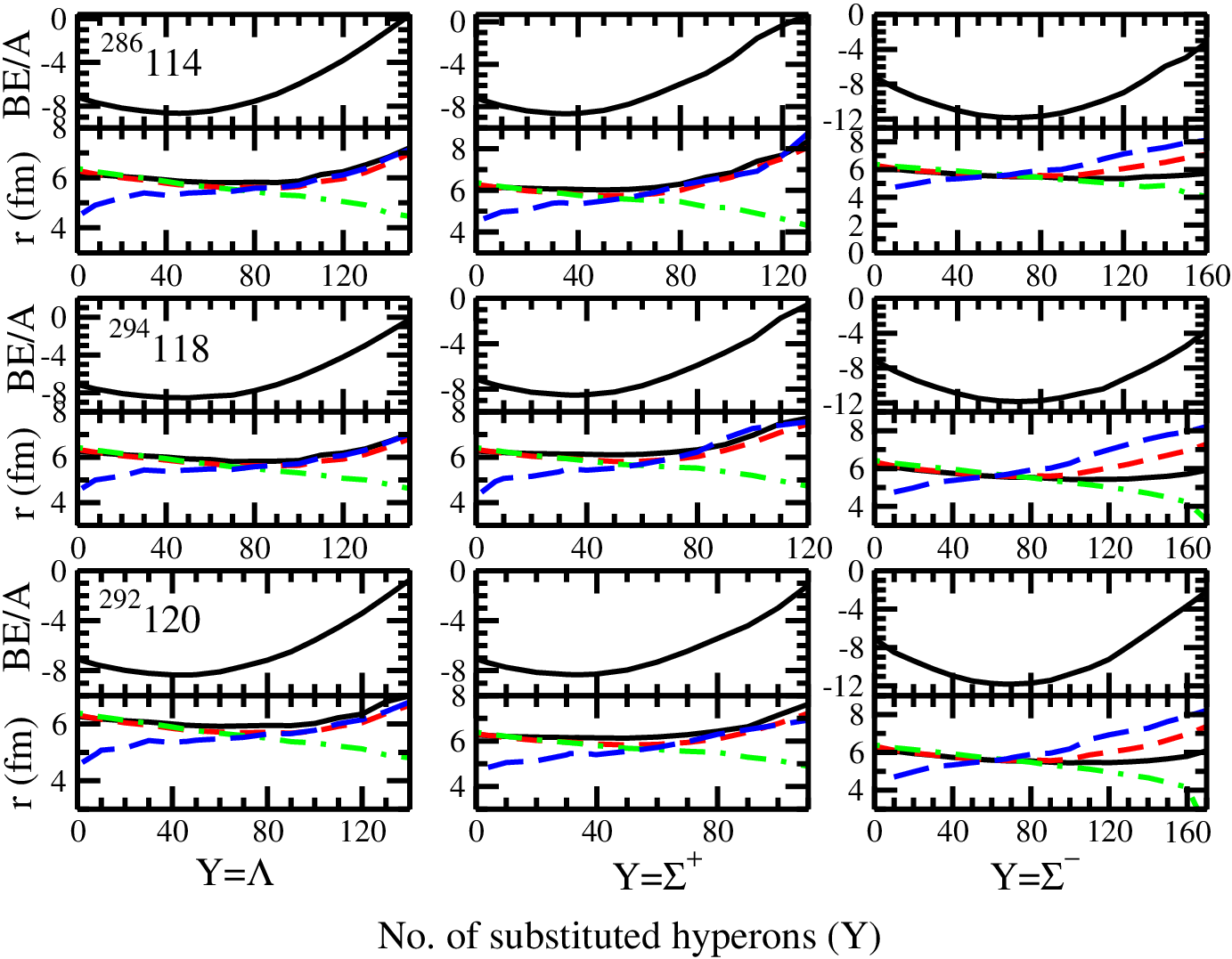}
\caption{\label{fig5} same as Fig. 1 but for $^{286}_{nY}$114, 
$^{294}_{nY}$118 and $^{292}_{nY}$120.}  
\end{figure}

\begin{figure}
\vspace{0.6cm}
\includegraphics[width=1.0\columnwidth]{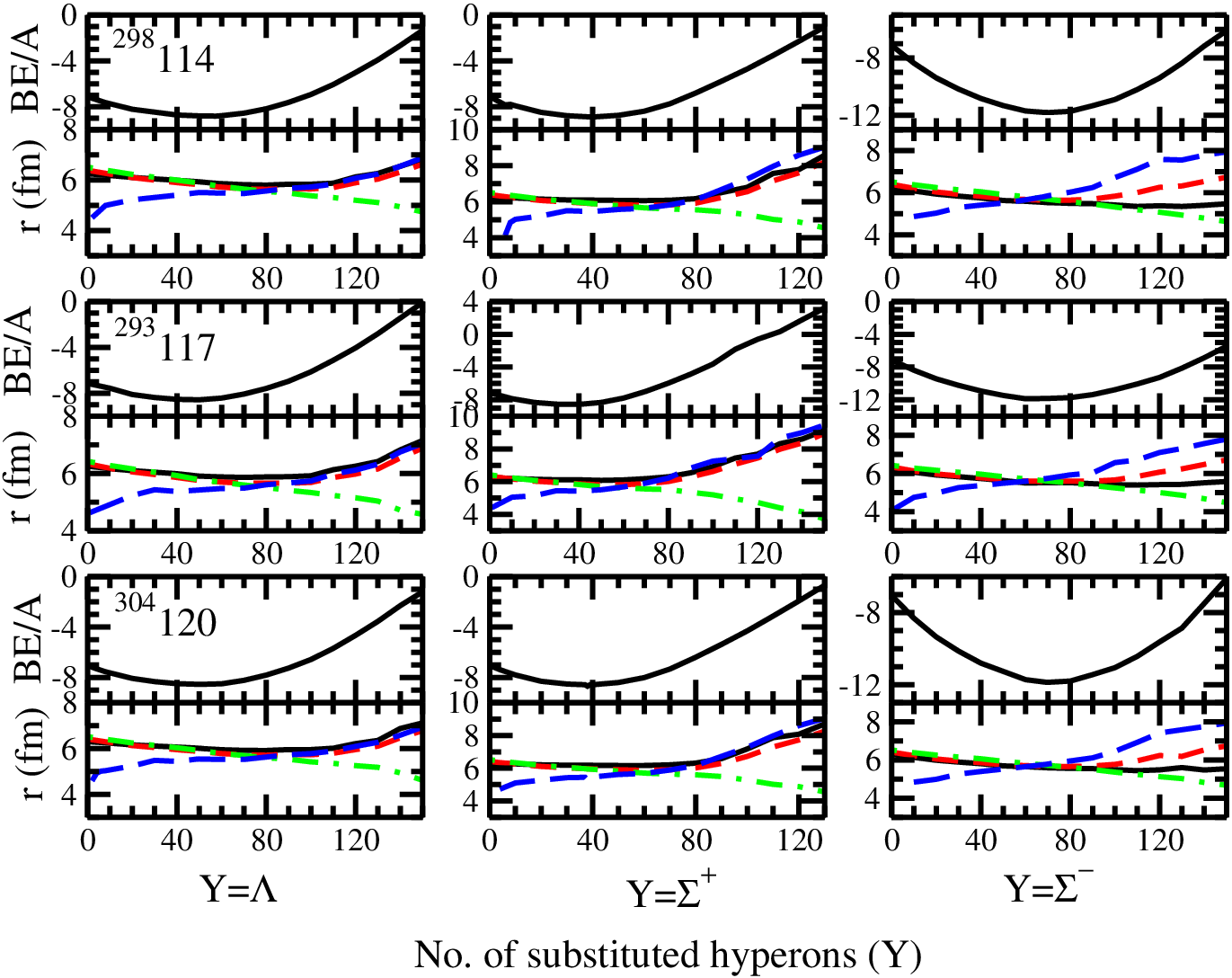}
\caption{\label{fig6} same as Fig. 1 but for $^{298}_{nY}$114, 
$^{293}_{nY}$117 and $^{304}_{nY}$120.}  
\end{figure}

\begin{figure}
\vspace{0.6cm}
\includegraphics[width=1.0\columnwidth]{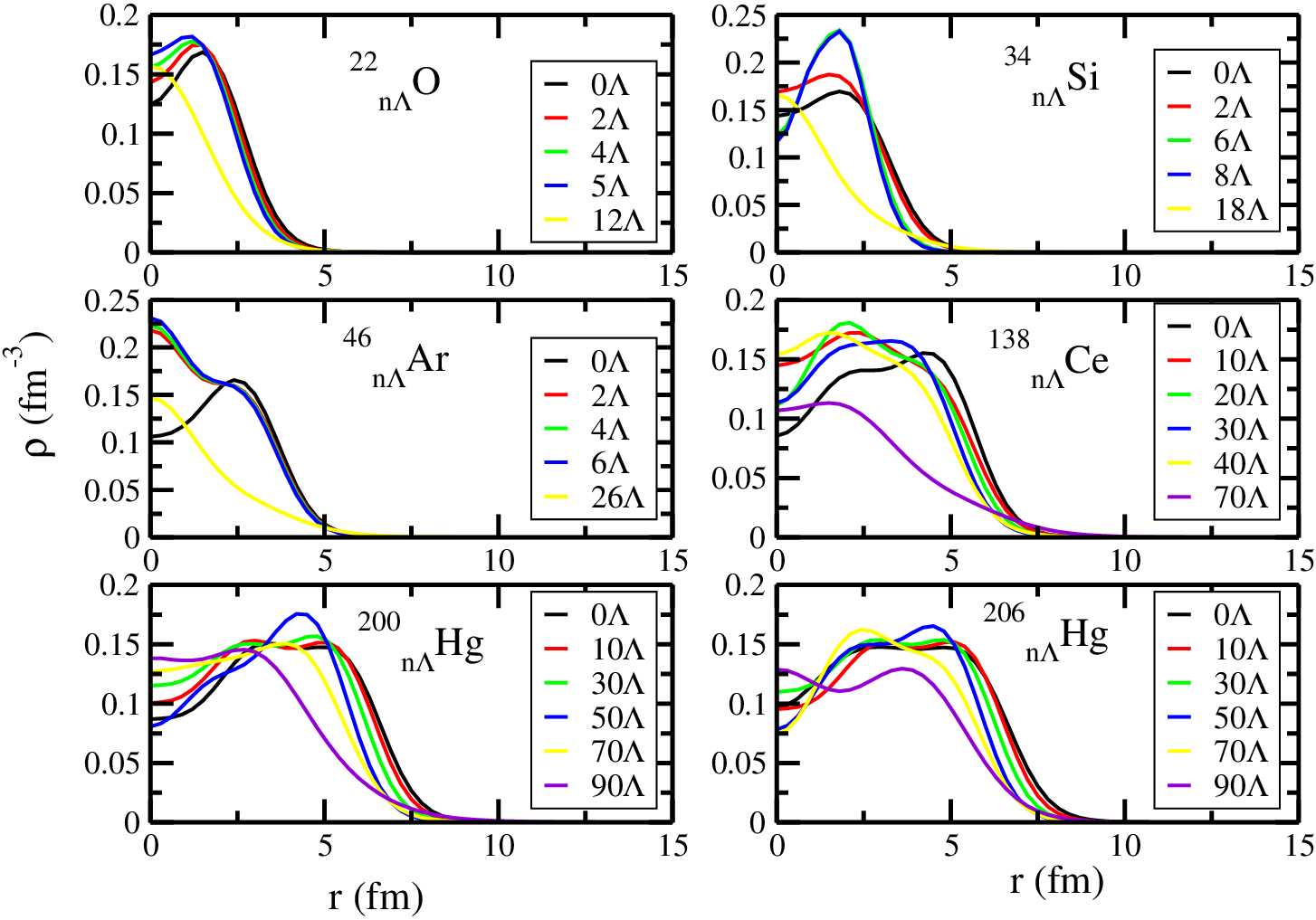}
\caption{\label{fig7} To see the depletion fraction 
especially for light bubble nuclei, the nucleon 
density is displayed for $^{22}_{n\Lambda}$O, 
$^{34}_{n\Lambda}$Si, $^{46}_{n\Lambda}$Ar, 
$^{138}_{n\Lambda}$Ce, $^{200}_{n\Lambda}$Hg 
and $^{206}_{n\Lambda}$Hg.}  
\end{figure}

\begin{figure}
\vspace{0.6cm}
\includegraphics[width=1.0\columnwidth]{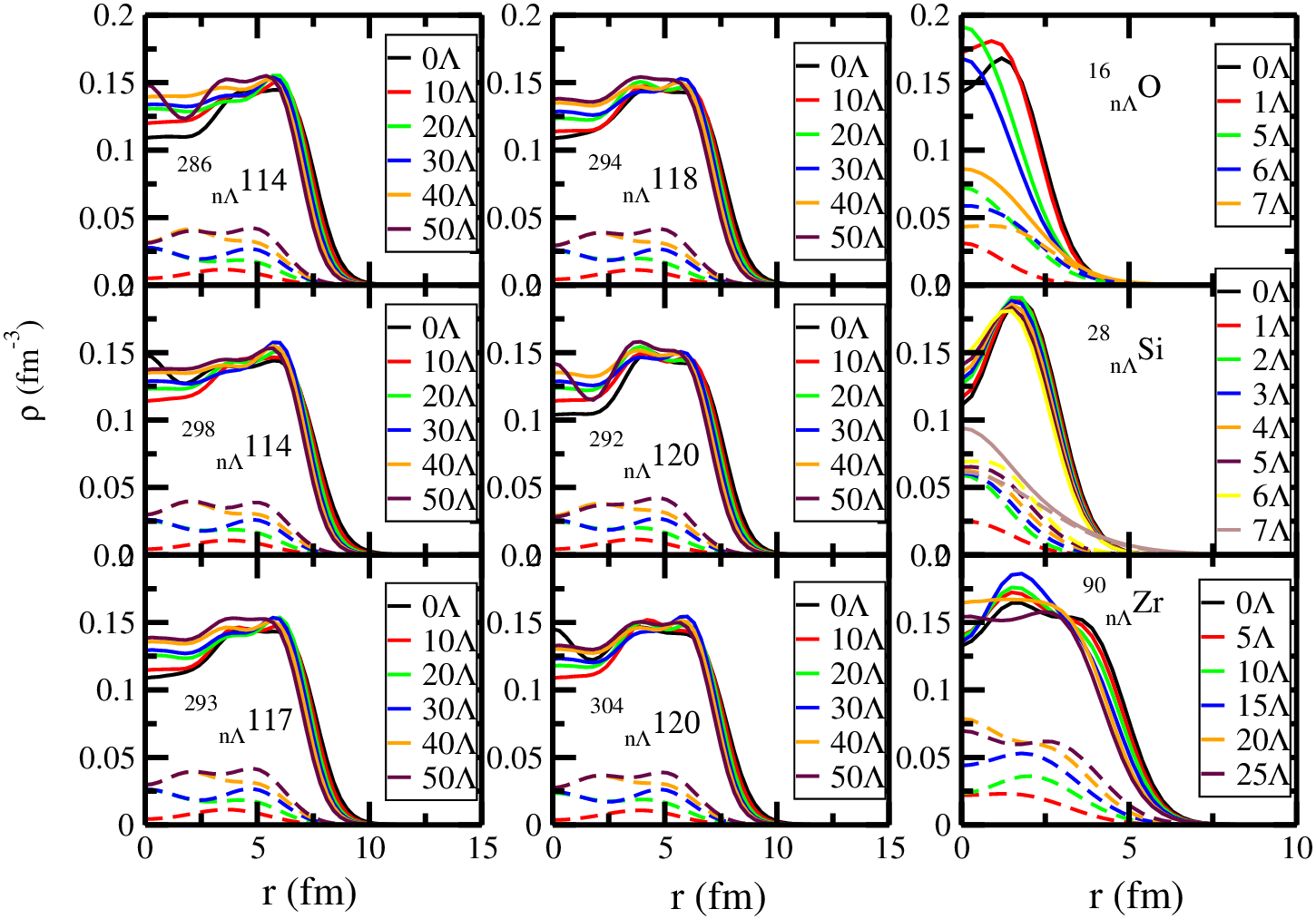}
\caption{\label{fig8} Nucleon and lambda densities for 
non-strange to multi-strange nuclei are displayed for 
$^{286}_{n\Lambda}$114, $^{298}_{n\Lambda}$114, 
$^{293}_{n\Lambda}$117, $^{294}_{n\Lambda}$118, 
$^{292}_{n\Lambda}$120, $^{304}_{n\Lambda}$120, 
$^{16}_{n\Lambda}$O, $^{28}_{n\Lambda}$Si and $^{90}_{n\Lambda}$Zr.
Solid lines represent the nucleon density and lambda 
densities represented by dashed lines.}  
\end{figure}

\begin{figure}
\vspace{0.6cm}
\includegraphics[width=1.0\columnwidth]{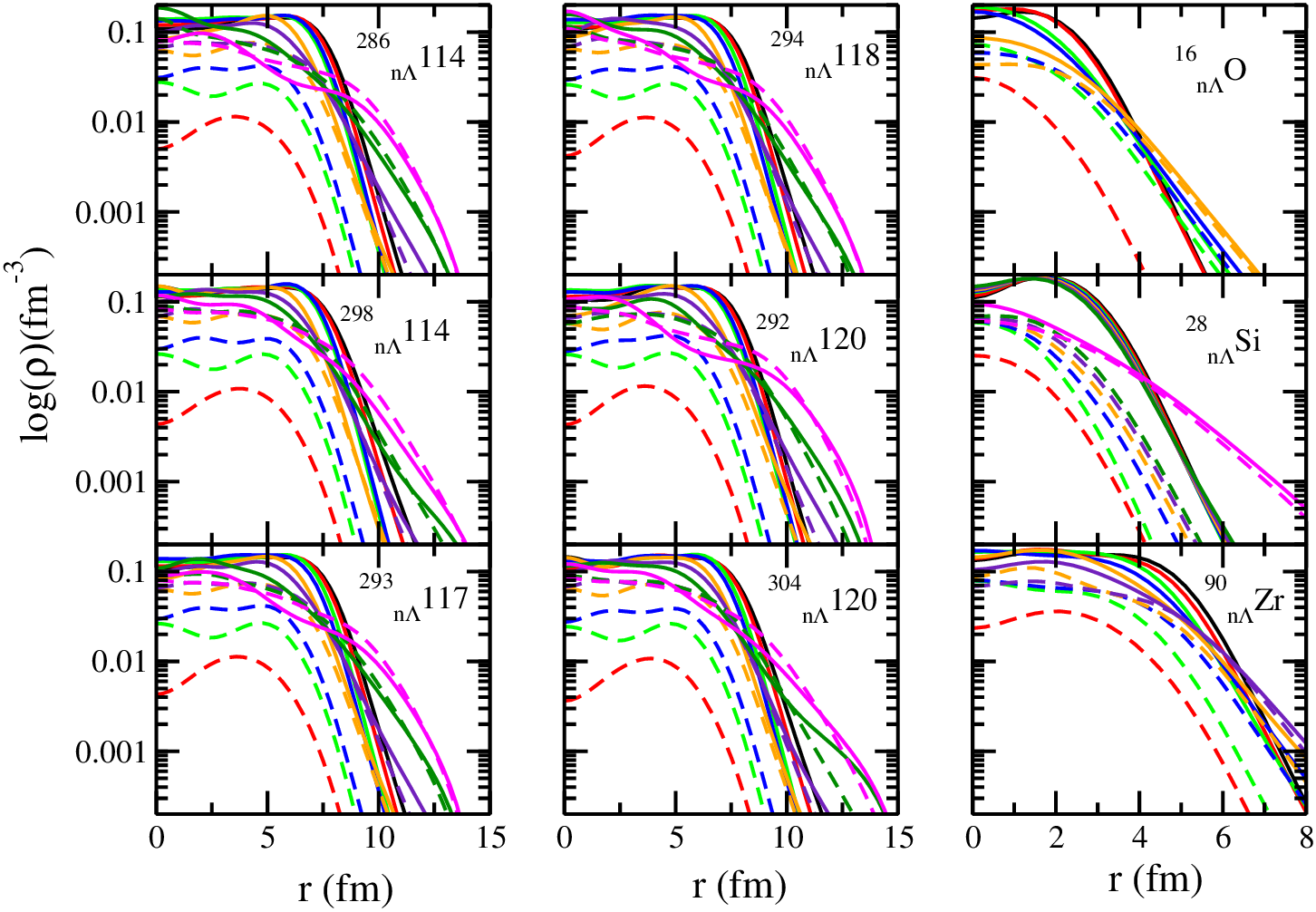}
\caption{\label{fig9} The nucleon and lambda densities 
of Figure 8 are plotted in logarithm scale to analyze 
the nucleon and lambda halo nature for 
$^{286}_{n\Lambda}$114, $^{298}_{n\Lambda}$114, 
$^{293}_{n\Lambda}$117, $^{294}_{n\Lambda}$118, 
$^{292}_{n\Lambda}$120, $^{304}_{n\Lambda}$120, 
$^{16}_{n\Lambda}$O, $^{28}_{n\Lambda}$Si and $^{90}_{n\Lambda}$Zr.  
In the same way as Fig 8, the nucleon and lambda 
densities are represented by solid and dashed lines, respectively.}  
\end{figure}

\begin{figure}
\vspace{0.6cm}
\includegraphics[width=1.0\columnwidth]{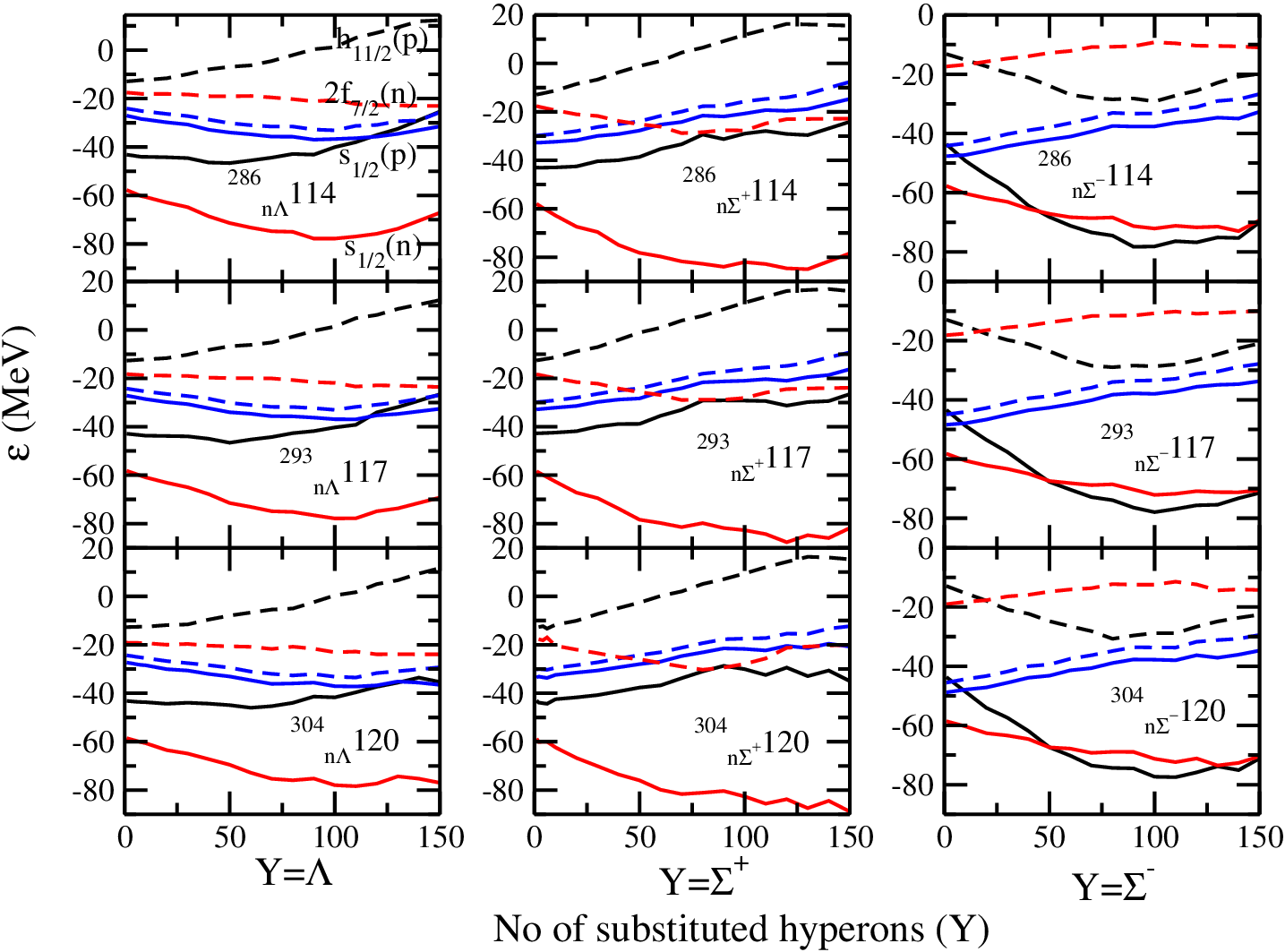}
\caption{\label{fig10}The first occupied and higher orbits 
of neutron and proton are shown for 
$^{286}_{n\Lambda}$114, $^{293}_{n\Lambda}$117, $^{304}_{n\Lambda}$120 and
$^{286}_{n\Sigma^+}$114, $^{293}_{n\Sigma^+}$117, $^{304}_{n\Sigma^+}$120 and 
$^{286}_{n\Sigma^-}$114, $^{293}_{n\Sigma^-}$117, $^{304}_{n\Sigma^-}$120 
as a function of substituted hyperons ($\Lambda$, $\Sigma^+$ and $\Sigma^-$) 
with replacing neutrons.
%First occupied level of neutron, proton and hyperon are represented 
%by solid lines with red, black and blue while the higher levels 
%are represented by dashed lines with their respective colors.}  
Solid lines with red, black and blue color represent the $1s_{1/2}$ 
level for neutron, proton and hyperon, respectively.
The higher neutron ($2f_{7/2}$), proton ($1h_{11/2}$) and hyperon 
($1p_{3/2}$) levels are represented by dashed line with red, black 
and blue color, respectively.}
\end{figure}

\begin{figure}
\vspace{0.6cm}
\includegraphics[width=1.0\columnwidth]{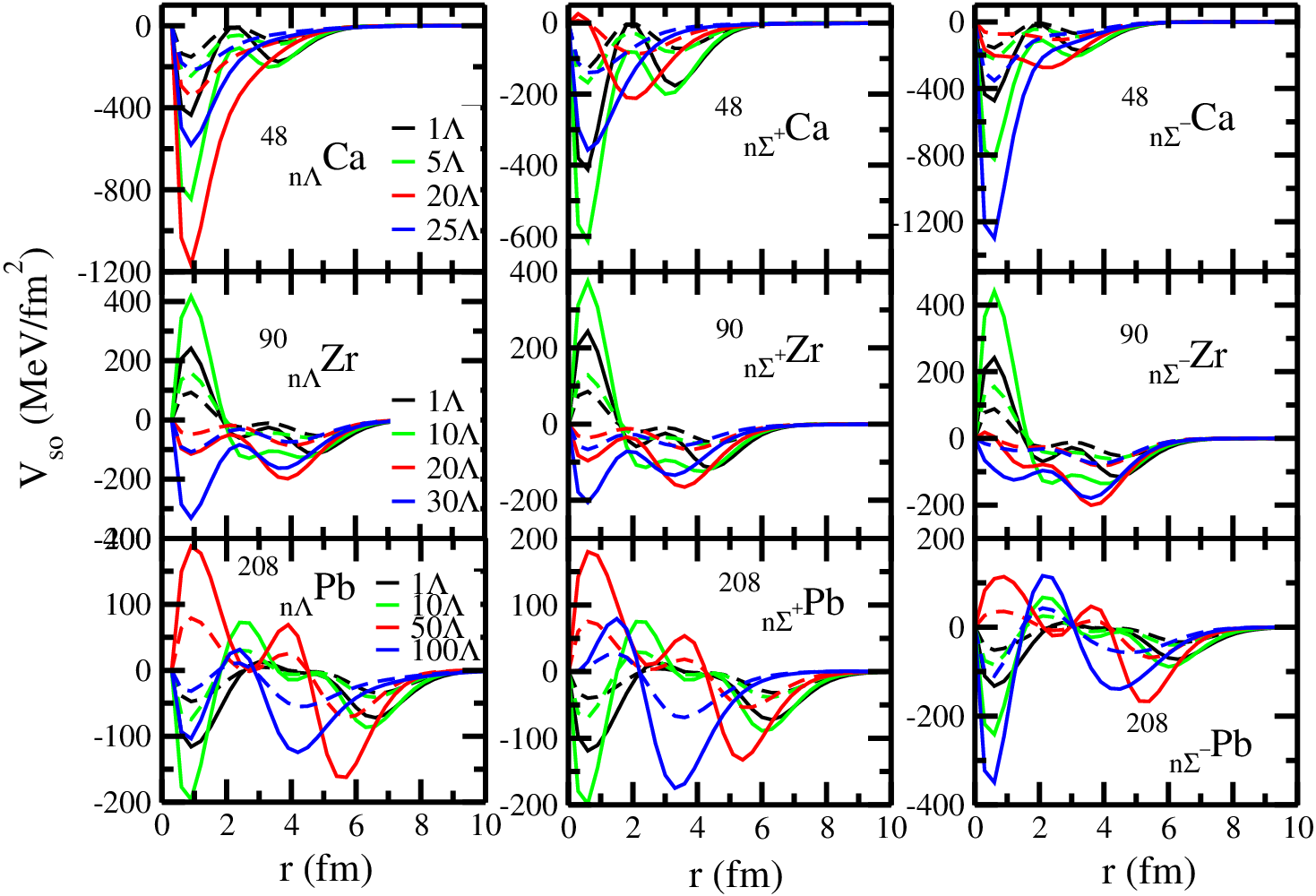}
\caption{\label{fig11} Radial dependence of spin-orbit potential 
for nucleon ($V_{so}^N$) and hyperons ($V_{so}^{\Lambda}$, 
$V_{so}^{\Sigma^+}$, $V_{so}^{\Sigma^-}$) are plotted for 
$^{48}_{n\Lambda}$Ca, $^{90}_{n\Lambda}$Zr, $^{208}_{n\Lambda}$Pb and
$^{48}_{n\Sigma^+}$Ca, $^{90}_{n\Sigma^+}$Zr, $^{208}_{n\Sigma^+}$Pb and 
$^{48}_{n\Sigma^-}$Ca, $^{90}_{n\Sigma^-}$Zr, $^{208}_{n\Sigma^-}$Pb.
Nucleon and hyperon spin-orbit potentials are represented by 
solid and dashed lines, respectively.}  
\end{figure}

\begin{figure}
\vspace{0.6cm}
\includegraphics[width=1.0\columnwidth]{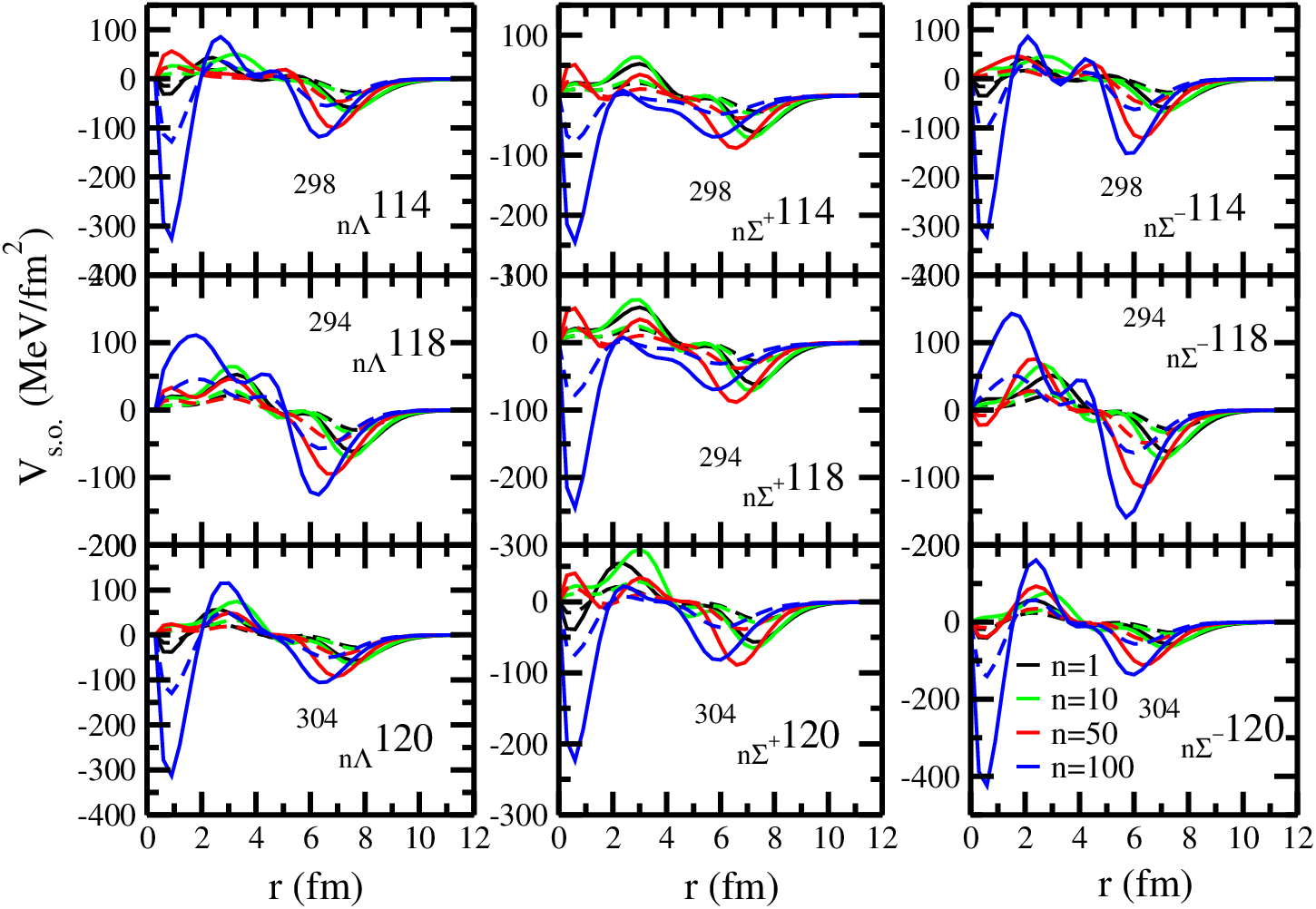}
\caption{\label{fig12} Same as Fig. 11 but for superheavy mass region 
$^{298}_{n\Lambda}$114, $^{294}_{n\Lambda}$118, $^{304}_{n\Lambda}$120 and 
$^{298}_{n\Sigma^+}$114, $^{294}_{n\Sigma^+}$118, $^{304}_{n\Sigma^+}$120 and 
$^{298}_{n\Sigma^-}$114, $^{294}_{n\Sigma^-}$118, $^{304}_{n\Sigma^-}$120.}  
\end{figure}

\begin{figure}
\vspace{0.6cm}
\includegraphics[width=1.0\columnwidth]{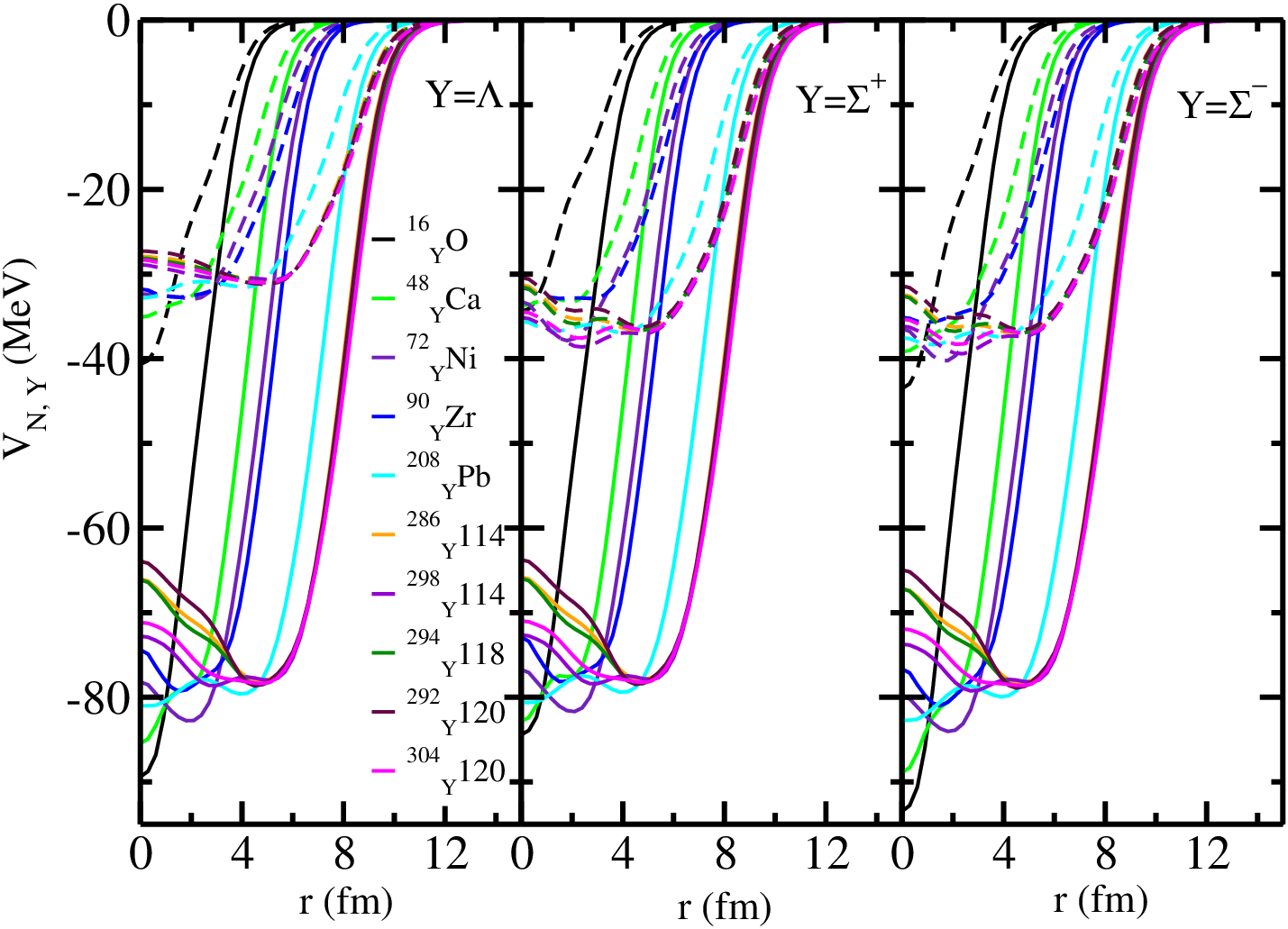}
\caption{\label{fig13} The neutron ($V_N$) and hyperon 
($V_Y=V_\Lambda, V_\Sigma^+, V_\Sigma^-$) mean 
potentials are plotted for $^{16}_Y$O, $^{48}_Y$Ca, 
$^{72}_Y$Ni, $^{90}_Y$Zr, $^{208}_Y$Pb, $^{286}_Y$114, 
$^{298}_Y$114, $^{294}_Y$118, $^{292}_Y$120, $^{304}_Y$120. 
Neutron and hyperon mean potentials are represented by solid 
and dashed lines, respectively.}  
\end{figure}

\begin{figure}
\vspace{0.6cm}
\includegraphics[width=1.0\columnwidth]{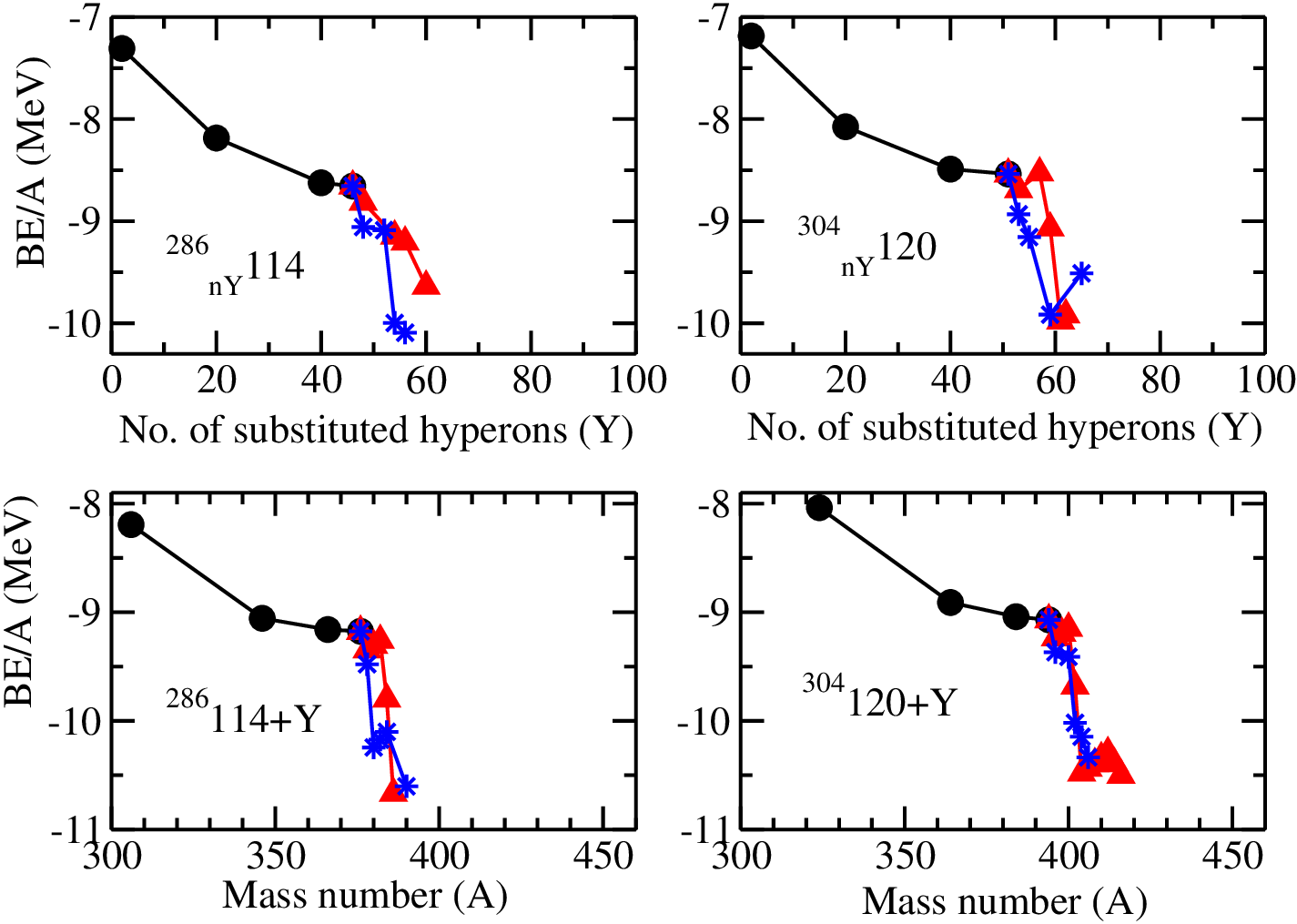}
\caption{\label{fig14} 
Binding energy per baryon (BE/A) as a function of 
substituted hyperons for $^{286}_{nY}$114 and 
$^{304}_{nY}$120 is displayed in upper portion of the figure.  
Binding energy per baryon (BE/A) as a function of mass 
number A for $^{286}$114+Y and $^{304}$120+Y, where 
Y=$\Lambda$+$\Sigma^+$,  $\Lambda$+$\Sigma^-$ is presented 
in lower part of the figure.
The solid circles in black color represent the binding 
for lambda, while the solid triangles in red color denote 
the binding for added $\Sigma^+$'s and the stars in blue 
color represent the binding by added $\Sigma^-$'s.}  
\end{figure}

\begin{table*}
\renewcommand{\arraystretch}{1.4}
\caption{\label{tab4}Total depletion fraction (D.F.) (in \%) 
as a measure of depletion of central density for nucleon 
distribution is listed for some selected bubble nuclei and 
their corresponding multi-strange hypernuclei.}
\begin{ruledtabular}
\begin{tabular}{lccccccccccc}
Nuclei&D.F. &Hypernuclei&D.F. &Hypernuclei&D.F. &Hypernuclei&D.F. &Hypernuclei&D.F. &Hypernuclei&D.F.  \\
\hline
$^{16}$O&15.11&$^{16}_{1\Lambda}$O&4.81 &$^{16}_{5\Lambda}$O&0.0  &$^{16}_{6\Lambda}$O&0.0  &$^{16}_{7\Lambda}$O&0.0  &$^{16}_{8\Lambda}$O&0.0  \\
$^{22}$O&26.22&$^{22}_{1\Lambda}$O&23.37&$^{22}_{2\Lambda}$O&17.94&$^{22}_{4\Lambda}$O&12.63&$^{22}_{5\Lambda}$O&8.42 &$^{22}_{12\Lambda}$O&0.0 \\
$^{28}$Si&41.45&$^{28}_{1\Lambda}$Si&38.59&$^{28}_{2\Lambda}$Si&33.75&$^{28}_{4\Lambda}$Si&25.77&$^{28}_{6\Lambda}$Si&18.35&$^{28}_{7\Lambda}$Si&0.0  \\
$^{34}$Si&15.28&$^{34}_{1\Lambda}$Si&12.57&$^{34}_{2\Lambda}$Si&9.82 &$^{34}_{6\Lambda}$Si&47.67&$^{34}_{8\Lambda}$Si&49.98&$^{34}_{18\Lambda}$Si&0.0 \\
$^{46}$Ar&36.03&$^{46}_{1\Lambda}$Ar&0.0  &$^{46}_{2\Lambda}$Ar&0.0  &$^{46}_{4\Lambda}$Ar&0.0  &$^{46}_{6\Lambda}$Ar&0.0  &$^{46}_{26\Lambda}$Ar&0.0 \\
$^{90}$Zr&19.29&$^{90}_{5\Lambda}$Zr&18.75&$^{90}_{10\Lambda}$Zr&21.03&$^{90}_{15\Lambda}$Zr&27.33&$^{90}_{20\Lambda}$Zr&1.62&$^{90}_{25\Lambda}$Zr&2.28 \\
$^{138}$Ce &44.91&$^{138}_{10\Lambda}$Ce&16.02&$^{138}_{20\Lambda}$Ce&38.35&$^{138}_{30\Lambda}$Ce&31.66&$^{138}_{40\Lambda}$Ce&10.73&$^{138}_{70\Lambda}$Ce&0.0  \\
$^{200}$Hg &42.35&$^{200}_{10\Lambda}$Hg&34.42&$^{200}_{30\Lambda}$Hg&23.47&$^{200}_{50\Lambda}$Hg&54.0 &$^{200}_{70\Lambda}$Hg&15.85&$^{200}_{90\Lambda}$Hg&5.08 \\
$^{206}$Hg &34.57&$^{206}_{10\Lambda}$Hg &37.05&$^{206}_{30\Lambda}$Hg &28.56&$^{206}_{50\Lambda}$Hg &52.69&$^{206}_{70\Lambda}$Hg &51.14&$^{206}_{90\Lambda}$Hg &0.0\\
$^{286}$114&24.62&$^{286}_{10\Lambda}$114&20.57&$^{286}_{20\Lambda}$114&16.09&$^{286}_{30\Lambda}$114&11.99&$^{286}_{40\Lambda}$114&8.54 &$^{286}_{50\Lambda}$114&4.19 \\
$^{298}$114&0.0  &$^{298}_{10\Lambda}$114&22.25&$^{298}_{20\Lambda}$114&22.96&$^{298}_{30\Lambda}$114&18.36&$^{298}_{40\Lambda}$114&13.04&$^{298}_{50\Lambda}$114&10.52\\
$^{293}$117&25.17&$^{293}_{10\Lambda}$117&22.44&$^{293}_{20\Lambda}$117&18.35&$^{293}_{30\Lambda}$117&15.83&$^{293}_{40\Lambda}$117&10.21&$^{293}_{50\Lambda}$117&9.66 \\
$^{294}$118&25.49&$^{294}_{10\Lambda}$118&22.93&$^{294}_{20\Lambda}$118&18.18&$^{294}_{30\Lambda}$118&16.02&$^{294}_{40\Lambda}$118&10.38&$^{294}_{50\Lambda}$118&10.70\\
$^{292}$120&30.85&$^{292}_{10\Lambda}$120&23.59&$^{292}_{20\Lambda}$120&19.96&$^{292}_{30\Lambda}$120&14.74&$^{292}_{40\Lambda}$120&10.96&$^{292}_{50\Lambda}$120&9.78 \\
$^{304}$120&0.89 &$^{304}_{10\Lambda}$120&28.37&$^{304}_{20\Lambda}$120&21.59&$^{304}_{30\Lambda}$120&20.40&$^{304}_{40\Lambda}$120&14.14&$^{304}_{50\Lambda}$120&7.54 \\
\end{tabular}
\end{ruledtabular}
\end{table*}

\subsection{Binding energies and radii}
The stability of multi-strange hypernuclei has been 
studied by introducing the lambda as well as sigma 
hyperon by replacing the neutrons.
Total and single particle energies for single- $\Lambda$ 
and $\Sigma$ hypernuclei are listed in the 
Tables~\ref{tab1}, \ref{tab2}, \ref{tab3}.
Total ($r_t$), charge ($r_{ch}$), neutron ($r_n$) and 
hyperon ($r_Y$) radii are also framed in these tables. 
The single particle energies for s- and p- states are compared 
with existing data for single$-\Lambda$ hypernuclei.
To check the stability of bound system with high strangeness 
in respect of injected hyperons, the binding energy per particle 
(BE/A) are plotted for light to superheavy hypernuclei as 
displayed in Figs.~\ref{fig1}$-$~\ref{fig6}.
In case of light mass region the binding energies are enhanced by 
introducing one or two hyperons and further it goes to reduce.
However, for heavy mass region, the BE/A increases with 
injection of large number of hyperons and form a more bound system 
than their normal counter parts, for example, inclusion of one lambda 
increases the binding of $^{16}_{n\Lambda}$O, the binding of 
$^{51}_{n\Lambda}$V increases up to the addition of 8 lambdas 
and the number of injected lambdas for superheavy hypernuclei, 
$^{304}_{n\Lambda}$120, goes to 51.
These numbers of lambda hyperons form a multi-strange bound 
system having maximum stability.
Not only lambda, we also look for the stability of multi-sigma hypernuclei.
In case of $\Sigma^+$, the maximum stability comes forward 
in comparison to $\Lambda$ and $\Sigma^-$ hypernuclei.
Because it has a repulsive sigma potential as well as enhance the 
repulsive Coulomb potential due to its positive charge and as a result 
the binding of multi-$\Sigma^+$ hypernuclei is less. 
Due to attractive Coulomb potential between $\Sigma^-$ and proton, 
the maximum stability for multi-$\Sigma^-$ hypernuclei 
is extended.
For example, injection of 51 $\Lambda$'s provide the maximum 
stability for $^{304}_{n\Lambda}$120 and 38 $\Sigma$'s for 
$^{304}_{n\Sigma^+}$120 while for this nuclei the number of 
injected $\Sigma^-$'s are 70 which produce the maximum stability.
By reducing the number of neutrons of nuclei, the neutron 
radius gradually decreases.
On contrary to this, and obviously the lambda, sigma radius 
increases with increasing the numbers of substituted hyperons 
($\Lambda$, $\Sigma^+$, $\Sigma^-$).
In some cases, the hyperon radius drastically increases by  
addition of hyperons, for example, in $^{16}_{nY}$O, 
$^{40}_{nY}$Ca, $^{51}_{nY}$V, $^{72}_{nY}$Ni and so on.
This behaviour of radii can be explained by internal shell 
structure by means of single particle energy levels.  
The total radius of the hypernuclei initially decreases 
and after certain limit it goes to increase. 
This behaviour of $r_t$ indicates that, by addition of hyperons 
the size of hypernucleus goes to shrink up to a certain limit 
and then extend the size.

\subsection{Density profile and single particle energies}
The nucleon distribution can be explained by density profile 
which has gross information about the structure of the nucleus.
Many of the bulk properties like binding energies, radii, single 
particle energies and density profile are affected by continuous 
injection of hyperons.
In this regard, we plot the lambda and nucleon density distribution 
for some light and superheavy hypernuclei as shown in Fig.~\ref{fig8}.
The nucleon and lambda density distributions are changed by 
addition of lambdas to normal nuclei.
The magnitude of lambda density increases due to increasing 
number of lambdas. 
By viewing the density profile, one can examine the most interesting 
feature of nuclei i.e. bubble structure, which is the measure of 
depletion of central density.
Anomalous behavior of density distribution is observed for bubble nucleus.
It shows a dip at the center and a hump nearby to it following a 
slow decreasing in density to zero at the surface.
Some of the interesting examples of bubble nuclei in superheavy 
region are $^{286}$114, $^{292}$120, $^{304}$120 as reported 
in Ref.~\cite{singh2013}.

The existence of bubble structure other than the 
spherical was first suggested by Wheeler~\cite{wheeler} 
and extensively studied by Wilson~\cite{wilson1946} 
however, later by Siemens and Bethe~\cite{siemens1967}. 
The explanation on occurrence of bubble nuclei have been 
made using several models like, independent particle 
and Hartree-Fock Model~\cite{wong1972,davies1972}.
One of the interesting thing in this context is, it is 
not confined to a particular region but have the 
possibility for light mass to superheavy region.
One may expect that, the mechanism behind the formation of bubble 
structure is the depopulation of s levels and as a result 
due to less bound lower s levels the radius increases and 
subsequently, central part of density decreases
~\cite{decharge2003,grasso2009}.
The bubble and semi-bubble structure for superheavy and 
hyperheavy mass region have been reported in 
Refs.~\cite{decharge1999,decharge2003}.
Not only depopulation of s-levels is responsible to make 
the hollow of central region but this may also be interpreted 
by s-d orbital inversion as discussed by Zhao {\it et al.} and 
E Khan {\it et al.}~\cite{zao2011,khan2008}.
In quantitative way, the amount of bubble effect can be 
measured by calculating depletion fraction (DF) using the 
relation~\cite{zao2011,khan2008};
\begin{equation*}
(DF)_\alpha=\frac{(\rho_{max})_\alpha-(\rho_{cen})_\alpha}{(\rho_{max})_\alpha}\times100 , 
\end{equation*}
where $\rho_{max}$, $\rho_{cen}$ represent the maximum and central 
density, respectively and $\alpha$ denotes the neutron and proton.
Not only syperheavy but some medium-heavy nuclei also have a good 
amount of depletion fraction as tabulated in Table~\ref{tab4}.
The depletion of central density is decreased or completely 
reduced by injection of hyperons to normal nuclei as shown 
in Fig.~\ref{fig7}, \ref{fig8}. 
For the case of $^{16}$O, the depletion fraction is 15.11, and this 
amount reaches to 4.81 for $^{16}_\Lambda$O and ultimately becomes 
zero for $^{16}_{5\Lambda}$O hypernucleus as given in Table~\ref{tab4}. 
In the same way, $^{286}$114 has a big amount of DF as 24.62 and 
this value reduced to 4.19 by injecting 50 lambdas to $^{286}$114 
with replacing neutrons.
In this way, we can say that the addition of lambdas to nuclei 
has the ability to remove the bubble structure partially or 
fully which is appeared in normal nuclei.
The effect of lambdas on density profile by means of effect on 
depletion fraction are shown in Figs.~\ref{fig7}, \ref{fig8}.
Some interesting bubble nuclei like as $^{22}$O, $^{34}$Si, 
$^{46}$Ar, $^{138}$Ce, $^{200}$Hg and $^{206}$Hg are  
considered as to reveal the effect of lambdas on density profile.
It is evident from Figs.~\ref{fig7}, \ref{fig8} and Table~\ref{tab4} 
that the injection of lambdas affects the DF partially or 
completely and as a result the amount of DF 
becomes very small or zero.
In the same time, the magnitude of nucleon density decreases 
because of decreasing number of neutrons.
For example, for $^{46}$Ar, the DF is 36 and this amount  
becomes zero by injection of 2 or more lambdas to the core nucleus.
This happens because of lambda particle resides at the centre of 
the nucleus and attracts the surrounding nucleon towards the 
centre and as a result central density becomes high and the 
hollow part of the centre is filled by partially or completely. 
This is one of the most important implication of hyperon 
to nuclei for removing the bubble nature of nuclei.
  
The other prospects of density profile is to analyze the 
halo nature of nucleon and lambda.
It is one of the interesting character of the nuclei which makes 
differ from the normal nuclei.
The halo nuclei has slowly decaying exponential tails extending 
beyond the size of the nucleus.
To examine the halo nature of nucleon and $\Lambda$ hyperon, 
we plot the density in logarithm scale.
%The halo nature is identified by the extended density in respect of r.
The halo nature is identified by wide space extension of density distribution.
It is evident from Fig.~\ref{fig9} that the $^{16}_{7\Lambda}$O, 
$^{28}_{7\Lambda}$Si, $^{90}_{40\Lambda}$Zr and $^{90}_{50\Lambda}$Zr 
show nucleon and lambda halo in light medium mass region.
In superheavy mass region, the hypernuclei 
$^{286}_{n\Lambda}114$, $^{298}_{n\Lambda}114$, $^{293}_{n\Lambda}117$, 
$^{294}_{n\Lambda}118$, $^{292}_{n\Lambda}120$, $^{304}_{n\Lambda}120$ 
with 140, 160 lambdas reveal the nucleon and lambda halo as 
shown in Fig.~\ref{fig9}.
In general, the addition of excessive lambdas corresponding to 
nuclei exhibit the neutron and lambda halo nature.
The reason is simple, the majority of lambda hyperons push 
out the nucleon towards periphery and formed nucleon halo.
The successive addition of hyperons by replacing the neutrons 
provide the deep binding of neutrons due to the symmetry energy. 
Because of successive addition of 
$\Lambda$ to the nuclei the $\Lambda$ separation energy 
reduces and weak binding of hyperon levels leads to the 
halo nature of hyperon as discussed in Ref.~\cite{lu2002}. 
The $\Lambda$ density is found to be maximum near the center 
from where it pushes the nucleons towards the low density 
regions both at the periphery and at the center. 
And due to this, the bubble structure is disappeared 
partially or completely and also $\Lambda$ and 
nucleon halo structure is seen.

Any kinds of change in a system is directly reflected from the 
single particle energy levels.  
To analyze the impact of hyperon on single particle energy levels, 
we plot the first and some higher filled neutron, proton 
and hyperon ($\Lambda$, $\Sigma^+$, $\Sigma^-$) levels as a 
function of substituted hyperons for superheavy region.
In case of lambda hypernuclei, the first filled neutron 
level goes deeper and first proton level also feels the attraction.
By decreasing the number of neutrons (injection of hyperons) the 
impact of Coulomb repulsion becomes higher and the upper 
proton level goes to unbound.
The deep binding of neutron level is because of decresing in 
symmetry energy due to substitution of neutron by $\Lambda$ 
as discussed in Ref.~\cite{jiang2006}.
For $\Sigma^+$ case, the attraction is less and Coulomb repulsion 
becomes high enough because of positive charge of $\Sigma^+$ hyperon.
On the other hand, due to $\Sigma^-$ hyperons the proton as 
well as neutron levels feel more bound.
Here in this case, the proton levels are bound enough than 
their neutron levels as shown in Fig.~\ref{fig10}.
The lambda energy levels feel to be constant or mild 
attractive with the function of injected lambdas. 
On contrary to this, the sigma levels go towards less 
binding which is clearly reflected from Fig.~\ref{fig10}.

\subsection{Spin-orbit and mean potentials}
In order to investigate the structural properties of nuclei, 
the spin-orbit interaction plays a significant role to produce 
the results in quantitative way.  
It is the beauty of RMF in which the spin-orbit splitting 
develops naturally by the exchange of scalar and vector 
mesons and this is not limited only for nuclei but exists  
in hypernuclei also~\cite{boguta1981,noble1980,vretenar1998}.
However, the spin-orbit potential in hypernuclei is weaker 
than their normal nuclear case as demonstrated 
in Refs.~\cite{brockmann1977,keil2000,ajimura2001}.
It is clearly shown in Figs.~\ref{fig11}, \ref{fig12} that 
the spin-orbit potential for hyperon is weaker than their 
normal counter parts and these results are consistent with 
existing predictions~\cite{brockmann1977,keil2000,ajimura2001}.
In this work, we study the spin-orbit potential for nucleons 
($V^N_{so}$) as well as hyperons 
($V^\Lambda_{so}$, $V^{\Sigma^+}_{so}$, $V^{\Sigma^-}_{so}$) 
of hypernuclei for different cases of hyperons.
The effect of large number of injected hyperons on 
spin-orbit potentials is significantly investigated and 
plotted in Figs.~\ref{fig11}, \ref{fig12} 
for medium to superheavy multi-strange hypernuclei.

The neutron ($V_N$), lambda ($V_\Lambda$) and sigma 
($V_\Sigma^+$, $V_\Sigma^-$) mean potentials are also 
investigated and plotted in Fig.~\ref{fig13} for light 
to superheavy hypernuclei.
The mean potential depth of lambda and sigma is found to be 30 MeV, 
which is in agreement of existing calculations~\cite{keil2000}.
The neutron potential depth for light hypernuclei lies approx 
80 MeV, which reduces to around 64 MeV for superheavy hypernuclei.
The shape of hyperon potential looks like to be same as 
neutron potential but only the amount of depth is different.
It is to be notice that the neutron potential looks like as 
V-shape type and shows the maximum depth around 78 MeV at r=4 fm, 
while this amount of depth is reduced to around 65 MeV at r=0 fm.
This is an indication of relatively low concentration 
of the particles at central region (r=0) which is the direct 
consequence of formation of bubble structure.
  
\subsection{Possible bound states of multi-strange systems}
The extension towards the systems of large strangeness has 
firstly been investigated in Refs.~\cite{rufa1990,
schaffner1994,schaffner1993,schaffner2000}.
Some of the theoretical calculations have been made and 
suggested the bound class of objects composed of neutrons, 
protons, $\Lambda$'s and $\Xi$'s for light-medium nuclei
~\cite{schaffner1993,schaffner1994}.
In this context, we deal the multi-strange system with 
other heavy hyperon.
The systems of high strangeness developed by addition of 
$\Lambda$'s have more bound nature than their normal 
nuclear counter parts.  
As we have seen in Figs.~\ref{fig1}$-$\ref{fig6}, where the BE/A 
increases for a certain limit of lambda particle and 
shows the maximum stability for a particular system.
Here, we try to suggest the stable systems having more 
heavier hyperons including lambdas for 
example, $\Sigma^+$ and $\Sigma^-$.
In this connection, we examine some combinations of 
n, p, $\Lambda$, $\Sigma^+$ and n, p, $\Lambda$, $\Sigma^-$
in the range of superheavy nuclei by two ways. 
In first one, we add the hyperons with replacing neutrons 
by staying the mass number as 
constant as shown in upper part of Fig.~\ref{fig14} while in 
second one, we add the hyperons in ordinary nuclei with 
increasing mass number and as a result the bound system with 
high strangeness is suggested.
We frame the bound state in such a way that after getting 
the maximum stability by addition of $\Lambda$'s, the $\Sigma^+$ 
and $\Sigma^-$ hyperons are injected which further increase
the binding of the system as shown in Fig.~\ref{fig14}.
The upper portion of Fig.~\ref{fig14} indicates that the 
more deeply bound than systems of same A are presented with 
replacing the neutrons by $\Lambda$'s and $\Sigma$'s.
The following possible combinations are presented as 
$n_\Lambda$=46, with $n_{\Sigma^+}$=2, 8, 10, 14 and 
$n_{\Sigma^-}$=2, 6, 8, 10 by maintaining 
A=286 for $^{286}_{nY}$114.
For $^{304}_{nY}$120 hypernuclei with maintaining the mass 
number A=304, the possible combinations are $n_\Lambda$=51, with 
$n_{\Sigma^+}$=2, 6, 8, 10, 11 and $n_{\Sigma^-}$=2, 4, 8, 14.
In other way, the possible combinations of nucleons and hyperons 
by adding the hyperons with increasing the mass 
number for $^{286}$114+Y are given as follow; $n_\Lambda$=90, with 
$n_{\Sigma^+}$=2, 4, 6, 8, 10 and $n_{\Sigma^-}$=2, 4, 6, 8, 14.
In the same way for $^{304}$120+Y, the combinations are given as 
$n_\Lambda$=90, with $n_{\Sigma^+}$=2, 4, 6, 8, 10, 14, 16, 18, 
20, 22 and $n_{\Sigma^-}$=2, 6, 8, 10, 12 which have increasing 
binding for a particular system.

\section{SUMMARY AND CONCLUSIONS}
It is really true that, the RMF produces quite excellent 
result not only for normal nuclei but hypernuclei also.
We demonstrate the various physical properties of hypernuclei 
within the RMF and also see the effects of 
successive addition of hyperons to nuclear bound system.
In which, we expose how does the binding energies, radii, density 
and single particle energies are affected by continuous 
injection of hyperons ($\Lambda, \Sigma^+, \Sigma^-$).

In this paper, we study the bulk properties as well as 
check the stability of hypernuclei within the RMF for a wide 
spectrum from light mass to superheavy region.
We investigate the binding energies, radii and single particle 
energies as a function of successive 
added $\Lambda$'s, $\Sigma^+$'s and $\Sigma^-$'s.
The stability of multi-strange hypernuclei is investigated 
as a function of added hyperons from light mass to superheavy region.
A variation in achieving the maximum stability for a particular system 
is observed for injection of different kinds of hyperons.
The study of bubble structure of the nuclei and the disappearance 
of bubble nature by addition of hyperons to normal nuclei is presented. 
The amount of depletion fraction of the nuclei is reduced 
by successive addition of hyperons.
Removing the bubble structure of nuclei by injection of lambdas 
is an important implication of strange baryons to ordinary nuclei. 
The nucleon and hyperon halo structure is also investigated.
The study of single-particle energy levels of $\Lambda$, 
$\Sigma^+$ and $\Sigma^-$hypernuclei is presented.
It is obsrved that the inclusion of hyperons affects the nucleon and 
hyperon spin-orbit potential significantly.
The neutron and hyperon ($\Lambda$, $\Sigma^+$, $\Sigma^-$) 
mean potentials are also displayed.

The bound class of n, p, $\Lambda$, $\Sigma^+$ and 
n, p, $\Lambda$, $\Sigma^-$ are suggested within the RMF.
The addition of lambda obviously increase the stability 
of the system but further addition of sigma hyperons 
leads to more stable system with increasing binding. 
The combinations of hyperons with nucleons for a particular 
system are suggested which will form the bound system with 
high strangeness and might be produced in heavy-ion reactions. 
The investigation on hypernuclei with high strangeness is quite 
welcome because these type of system might be produced 
in high-energy heavy ion reactions near the future.
These types of study of hypernuclei with multiple strangeness 
will provide us a basic input for neutron as well as hyperon 
star studies, which are the body of current interest.

\section*{ACKNOWLEDGMENTS}
One of the author (MI) would like to acknowledge the hospitality 
provided by Institute of Physics, Bhubaneswar during the work.

\newpage
%%%%%%%%%%%%%%%%%%%%%%%%%%%%%%%%%%%%%%%%%%%%%%%%%%%%%%%%%%%%%%%%

\end{document}